\documentclass[preprint]{vldb}
 \usepackage{cite}
 \usepackage{amssymb}
 \usepackage{amsfonts}
 \usepackage{tikz,pgfplots}
 \usepackage{amsmath}
 \usepackage{subfigure}
 \usepackage{mathtools}
 \usepackage{balance}
 \usepackage{xspace}
\usepackage{algorithm}
\usepackage{algorithmicx}
\usepackage{algpseudocode}
\usetikzlibrary{positioning}
 \pgfplotsset{compat=1.14}

\newcommand{\protocol}{SeeMoRe\xspace}
\newcommand{\req}{\textsf{request}\xspace}
\newcommand{\REQ}{\textsf{REQUEST}\xspace}
\newcommand{\zero}{\textsf{pre-prepare}\xspace}
\newcommand{\one}{\textsf{prepare}\xspace}
\newcommand{\ONE}{\textsf{PREPARE}\xspace}
\newcommand{\two}{\textsf{accept}\xspace}
\newcommand{\TWO}{\textsf{ACCEPT}\xspace}
\newcommand{\three}{\textsf{commit}\xspace}
\newcommand{\THREE}{\textsf{COMMIT}\xspace}
\newcommand{\reply}{\textsf{reply}\xspace}
\newcommand{\REPLY}{\textsf{REPLY}\xspace}
\newcommand{\inform}{\textsf{inform}\xspace}
\newcommand{\INFORM}{\textsf{INFORM}\xspace}
\newcommand{\vchange}{\textsf{view-change}\xspace}
\newcommand{\VCHANGE}{\textsf{VIEW-CHANGE}\xspace}
\newcommand{\checkp}{\textsf{checkpoint}\xspace}
\newcommand{\CHECKP}{\textsf{CHECKPOINT}\xspace}
\newcommand{\newv}{\textsf{new-view}\xspace}
\newcommand{\NEWV}{\textsf{NEW-VIEW}\xspace}

\newcommand{\MCHANGE}{\textsf{MODE-CHANGE}\xspace}
\newcommand{\calo}{\mathcal{O}}

 \def\caP{{\cal P}}
 
 \algrenewcommand\alglinenumber[1]{\scriptsize #1:}

\makeatletter
\def\@copyrightspace{\relax}
\makeatother

\begin{document}

\title{SeeMoRe: A Fault-Tolerant Protocol for Hybrid Cloud Environments}

 \author{
 \alignauthor
 Mohammad Javad Amiri \qquad Sujaya Maiyya \qquad Divyakant Agrawal \qquad Amr El Abbadi\\
      \affaddr{Department of Computer Science, University of California Santa Barbara}\\
      \affaddr{Santa Barbara, California}\\
      \email{\{amiri, sujaya\_maiyya, agrawal, amr\}@cs.ucsb.edu}
 }

\maketitle

\begin{abstract}
Large scale data management systems utilize State Machine Replication to
provide fault tolerance and to enhance performance. Fault-tolerant protocols are extensively
used in the distributed database infrastructure of large enterprises such as Google, Amazon, and Facebook,
as well as permissioned blockchain systems like IBM's Hyperledger Fabric.
However, and in spite of years of intensive research, 
existing fault-tolerant protocols
do not adequately address all the characteristics of
distributed system applications. 
In particular, hybrid cloud environments consisting of private and public clouds
are widely used by enterprises.
However,
fault-tolerant protocols have not been adapted for
such environments.
In this paper, we introduce {\em \protocol},
a hybrid State Machine Replication protocol to
handle both crash and malicious failures in a public/private cloud environment.
\protocol considers a private cloud consisting of non-malicious  nodes (either correct or crash) and
a public cloud with both Byzantine faulty and correct nodes.
\protocol has three different modes which can be used depending on
the private cloud load and
the communication latency between the public and the private cloud.
We also introduce a dynamic mode switching technique to transition from one mode to another.
Furthermore, we evaluate \protocol
using a series of benchmarks.
The experiments reveal that \protocol's performance is close to the state of the art 
crash fault-tolerant protocols
while tolerating malicious failures.
\end{abstract}
\section{Introduction \label{sec:intro}}

Today's enterprises mostly rely on cloud
storage to run their business applications.
Cloud computing has many benefits in terms of cost savings, scalability,
and easy access \cite{zhang2010cloud}.
However, storing data on a single cloud may reduce
robustness and performance \cite{cachin2009trusting,dobre2014hybris,gunawi2016does}.
Robustness is the ability to ensure availability (liveness) and
one-copy semantics (safety) despite failures, while
performance deals with the response time of requests (latency) and 
the number of processed requests per time unit 
(throughput) \cite{aublin2015next}.

Fault-tolerant protocols are designed to satisfy both
robustness and performance concerns using State Machine
Replication (SMR) \cite{lamport1978time} techniques.
SMR regulates the
deterministic execution of client requests on
multiple copies of a server, called replicas, such that
every non-faulty replica must execute every request in 
the same order
\cite{schneider1990implementing}\cite{lamport1978time}.

Large scale data management systems utilize SMR to
provide fault tolerance and to increase the performance
of the system. Fault-tolerant protocols are extensively
used in distributed databases such as Google's
Spanner~\cite{corbett2013spanner}, Amazon's
Dynamo \cite{decandia2007dynamo}, and Facebook's Tao \cite{bronson2013tao}, thus highlighting
the critical role of SMR in data management. 
SMR is also the core component in the more
recently developed, highly popular set of technologies
-- \textit{Blockchain}.
In particular, permissioned blockchain systems
extensively use fault-tolerant protocols to establish 
consensus
on the order of transactions 
between a set of known,
identified nodes that do not fully trust each other.

While large enterprises might have their own Geo-replicated fault-tolerant cloud storage around the world,
smaller enterprises may only have a local private cloud that is lacking in
resources to guarantee fault tolerance.
One solution is to store all the data on third-party public cloud providers
\cite{armbrust2010view,wu2013spanstore,buasescu2012robust}.
Public clouds provide several advantages like elasticity and durability,
but they often suffer from security concerns,
e.g., malicious attacks \cite{alzain2011mcdb}.
Whereas private clouds are considered more secure but may
not provide sufficient elasticity and durability.
The trustworthiness of a private cloud allows an enterprise to build services that can
utilize crash fault-tolerant protocols,
i.e., protocols that make progress when a bounded number of 
replicas only fail in a benign manner, for example by either crashing or being unresponsive. 
But due to lack of private resources,
if a third-party public cloud is used, the nodes of the public cloud \textit{may} behave maliciously, 
in which case a more robust fault-tolerant protocol is needed that allows the system to continue operating 
correctly, even when some replicas exhibit arbitrary, possibly malicious behavior.
Current Byzantine fault-tolerant protocols (e.g., PBFT~\cite{castro1999practical})
introduce significant communication and latency overheads in order to tolerate failures
since they consider all failures as malicious.

An alternative solution to storing all the data in public cloud is to use
a hybrid cloud storage system consisting of both private and public clouds \cite{dobre2014hybris}.
In a hybrid cloud, the nodes in the private cloud are trusted and may crash but do not behave maliciously
whereas the nodes in the public cloud(s) might be malicious.
Hybrid clouds address the {\em security} concerns of using {\em only} public clouds by
giving enterprises the ability to still use their private clouds.
In addition, storing data on multiple clouds is more {\em reliable}, e.g.,
if a cloud outage happens, the system might still be able to process requests.
Moreover, while the small private cloud may represent a {\em scalability} bottleneck,
the system can rent as many servers as required wants from the public clouds.
The benefits of a hybrid cloud
necessitates designing new protocols that can leverage the trust of private clouds and the
scalability of public clouds.

Despite years of intensive research, existing fault-tolerant protocols
do not adequately address all the characteristics of hybrid cloud environments. 
On one hand, the existing Byzantine fault-tolerant protocols
\cite{castro1999practical,martin2006fast,kotla2007zyzzyva,wood2011zz,kapitza2012cheapbft,aublin2015next,liu2016xft}
do not distinguish between crash and malicious failures,
and consider all failures as malicious, 
thus incurring a higher cost in terms of performance. 
On the other hand, the hybrid protocols 
\cite{serafini2010scrooge,clement2009upright} that have been 
designed to tolerate both
crash and malicious failures,
make no assumption on {\em where} the crash or malicious failures 
may occur.
As a result using these protocols in a hybrid cloud environment, 
where all machines in the private cloud are known to be trusted 
while machines in the public cloud could be 
compromised and hence malicious, results in
an unnecessary performance overhead.

A hybrid fault tolerant protocol can be beneficial in
a distributed database such as Spanner~\cite{corbett2013spanner}. 
Spanner consists of a layered architecture
where transaction management is oblivious to the underlying 
fault-tolerant protocol used. If a small enterprise wants to 
deploy Spanner in a hybrid cloud, it can plug a hybrid
fault-tolerant protocol instead of the crash fault-tolerant
protocol used in Spanner to achieve more reliable replication
in this heterogeneous cloud setting.
Permissioned blockchain systems can also take advantage of a
hybrid protocol.
For example, in IBM's Hyperledger Fabric~\cite{androulaki2018hyperledger} the fault-tolerant (consensus) protocol is pluggable.
Hence, if some nodes are trusted but not all, Fabric can benefit from a hybrid fault-tolerant protocol.

In this paper, we present {\em \protocol}\footnote{ \protocol is derived from Seemorq, a benevolent, mythical bird in Persian mythology which appears as a peacock
with the head of a dog and the claws of a lion.
Seemorq in Persian literature also refers to a group of birds who flew together to achieve a common goal.}:
a State Machine Replication protocol that leverages the localization of
crash and malicious failures in a {\em hybrid cloud environment}.
\protocol considers a private cloud consisting of trusted replicas, a subset of which may fail-stop, and
a public cloud where a subset of the replicas may behave maliciously.
\protocol takes explicit advantage of this knowledge to improve performance by reducing
the number of communication phases and messages exchanged and/or the number of required replicas.
\protocol has three different modes of operation which can be used depending on
the load on the private cloud, and the latency between the public and the private cloud.
We also introduce a dynamic and elastic technique to transition from one mode to another.

A key contribution of this paper is to show how being aware of {\em where} different types of failures (crash and malicious) may occur in hybrid cloud environments, results in
designing more efficient protocols.
In particular, this paper makes the following contributions:

\begin{itemize}
\item A model for hybrid cloud environments is presented which can be used by enterprises
that do not have enough servers in their trusted private cloud to run fault-tolerant protocols and
gives them the option of renting from untrusted public clouds.

\item \protocol, a hybrid protocol that tolerates both crash and malicious failures,
is developed in three different modes.
Being aware of
where the crash faults may occur and where the malicious faults can occur
results in reducing
the number of communication phases, messages exchanged and/or required replicas.
In addition, a technique to dynamically switch between different modes
of \protocol is presented.

\end{itemize}

The rest of this paper is organized as follows.
Section~\ref{sec:related} presents related work.
The system model is introduced in Section~\ref{sec:model}.
Section~\ref{sec:public} presents a method to compute the required number of replicas from a public cloud.
The design of \protocol is proposed in Section~\ref{sec:alg}, elaborating on the three modes, as well as the dynamic switching of modes.
Section~\ref{sec:eval} shows the performance evaluation, and
Section~\ref{sec:conc} concludes the paper.
\section{Related Work}\label{sec:related}

State machine replication (SMR) is a technique for implementing a fault-tolerant service
by replicating servers \cite{lamport1978time}.
Several approaches \cite{schneider1990implementing,lamport2001paxos,ongaro2014search}
generalize SMR to support
crash failures among which Paxos is the most well-known \cite{lamport2001paxos}.
Paxos guarantees safety in an asynchronous network using $2f{+}1$ processors
despite the simultaneous crash failure of any $f$ processors.
Many protocols are proposed to either
reduce the number of phases, e.g.,
Multi-Paxos which assumes the leader is relatively stable or
Fast Paxos \cite{lamport2006fast} and  Brasileiro et al. \cite{brasileiro2001consensus} which add $f$ more replicas,
or reduce the number of replicas, e.g., Cheap Paxos \cite{lamport2004cheap} which
tolerates $f$ failures with $f{+}1$ active and $f$ passive processors.

Byzantine fault tolerance refers to
servers that behave arbitrarily
after the seminal work by Lamport, et al. \cite{lamport1982byzantine}.
Practical Byzantine fault tolerance protocol (PBFT) \cite{castro1999practical}
is one of the first and probably the most instructive state machine replication protocol
to deal with Byzantine failures.
Although practical, the cost of implementing PBFT is quite high,
requiring at least $3f+1$ replicas, $3$ communication phases, and
a quadratic number of messages in terms of the number of replicas.
Thus, numerous approaches have been proposed to explore
a spectrum of trade-offs between the number of phases/messages (latency),
number of processors,
the activity level of participants (replicas and clients), and
types of failures.

On latency, FaB \cite{martin2006fast} and Bosco \cite{song2008bosco}
reduce the communication phases by adding more replicas.
Speculative protocols,
e.g., Zyzzyva \cite{kotla2007zyzzyva}, HQ \cite{cowling2006hq}, and
Q/U \cite{abd2005fault}, also reduce the communication by
executing requests without running any agreement between replicas
and optimistically rely on clients to detect inconsistencies between replicas.

To reduce the number of replicas,
some approaches rely on a trusted component
(a counter in A2M-PBFT-EA \cite{chun2007attested},
MinBFT \cite{veronese2013efficient} and,
EBAWA \cite{veronese2010ebawa},
a hypervisor \cite{reiser2007hypervisor},
or a whole operating-system instance \cite{correia2004tolerate})
that prevents a faulty replica from sending conflicting
(i.e., asymmetric) messages to different
replicas without being detected.

In addition, optimistic approaches reduce the required number of replicas
during the normal-case operation
by either utilizing the Cheap Paxos \cite{lamport2004cheap} solution and
keeping $f$ replicas in a passive mode
(REPBFT \cite{distler2016resource}), or
by separating agreement from
execution \cite{yin2003separating}.
In ZZ \cite{wood2011zz} both passive replicas and separating agreement
from execution are employed.
Note that all these approaches need $3f+1$ replicas upon occurrence of failures.
REMINBFT \cite{distler2016resource},
SPARE \cite{distler2011spare}, and
CheapBFT \cite{kapitza2012cheapbft}
use a trusted component to reduce the network size to $2f+1$ and then
keep $f$ of those replicas passive during the normal-case operation.
In contrast to optimistic approaches, robust protocols
(Prime \cite{amir2011prime}, Aardvark \cite{clement2009making},
Spinning \cite{veronese2009spin}, RBFT \cite{aublin2013rbft})
consider the system to be under attack by a very strong adversary and
try to enhance the performance of the protocol during periods of failures.

In this paper we focus on Hybrid fault tolerance.
Such consensus with multiple failure modes were initially addressed in
synchronous protocols \cite{thambidurai1988interactive,meyer1991consensus,kieckhafer1994reaching,siu1996note}.
Many recent efforts have focused on partial synchrony,
a technique that defines a
threshold on the number of slow (partitioned) processes.
Let $m$, $c$, and $s$ denote
the number of malicious, crash, and slow servers respectively,
VFT \cite{porto2015visigoth} and XFT \cite{liu2016xft} require
 $2m+c+min((m{+}c),s)+1$ and  $2(m+c+s)+1$ servers respectively, and
SBFT \cite{gueta2018sbft} needs $3m+2c+1$  
servers as the minimum size of the network from which $3m+c+1$ are participating
in each quorum.
VFT is very similar to PBFT regarding the number of phases and massage exchanges. 
XFT however, optimistically assumes that an adversary cannot fully control
the Byzantine nodes and as a result,
reduces the phases of communication and message exchanges.
SBFT also reduces the number of message exchanges by assuming the adversary controls
only the crash failures.

Finally, Scrooge \cite{serafini2010scrooge} and UpRight \cite{clement2009upright}
are two asynchronous hybrid approaches that use optimistic solutions.
Scrooge \cite{serafini2010scrooge} uses a speculative technique
to reduce the latency in the presence of $4m+2c$ replicas.
UpRight \cite{clement2009upright}, which is the closest protocol to SeeMoRe,
requires $3m+2c+1$ nodes as the minimum network size from which $2m+c+1$ are required to participate
in each communication quorum.
In addition, UpRight utilizes the agreement routines of
PBFT \cite{castro1999practical}, Aardvark \cite{clement2009making}, and Zyzzyva \cite{kotla2007zyzzyva}
and similar to \cite{yin2003separating},
separates agreement from execution.
However, UpRight is not aware which nodes may crash and which may be malicious, therefore,
does not take advantage of this knowledge by placing particular processes
executing specific protocol roles on crash-only or malicious sites.
On the other hand, SeeMoRe knows where the crash or malicious faults may occur,
thus, it either reduces the number of communication phases and message exchanges
by placing the primary in the crash-only private cloud,
or decreases the number of required nodes by placing the primary in the untrusted public cloud.

Storing data on multiple clouds to enhance fault tolerance is addressed for
both crash
(ICStore \cite{buasescu2012robust}, SPANStore \cite{wu2013spanstore}) and
malicious
(DepSky \cite{bessani2013depsky}, SCFS \cite{bessani2014scfs})
failures which rely on $2f+1$ and $3f+1$ servers respectively.
DAPCC \cite{wang2011achieving}
assumes a synchronous cloud environment and
solves the consensus in a dual failure mode
with $n\geq \lfloor(n-1)/3\rfloor +2m+c$ nodes
(similar to \cite{siu1996note}).
However, DAPCC needs $\lfloor(n-1)/3\rfloor+1$ rounds of communication.
Hypris \cite{dobre2014hybris} on the other hand
reduces the number of required servers to $2f+1$
($f+1$ when the system is synchronous and no faults happen)
by keeping the metadata in a private cloud assumed to be partially synchronous.

\section{System Model}\label{sec:model}

In this section, we introduce the system model
wherein an application layer, 
such as a distributed database management system,
relies on a replication service to store copies of data
across a cloud environment consisting of private and public 
clouds. The replication service aims to replicate the
data across some trusted and some untrusted servers. The
replicas can be geo-distributed to provide low data access 
latency to clients across the globe or they can be 
geographically confined to tolerate both
crash or malicious failures.
Such a replication service can use \protocol 
and we specify the assumptions on
which \protocol is built in this section. 

\subsection{Basic Assumptions}
We consider a hybrid failure model that admits both
crash and malicious failures
where crash failures may occur in the private cloud and malicious failures 
may {\em only} occur in the public cloud. Note that a 
malicious failure can encompass a crash failure but since the
trust assumptions are low, we do not distinguish between a 
crash or a malicious failure in the public cloud.
In a crash failure model, replicas operate at arbitrary speed,
may fail by stopping, and may restart, however they may not collude, lie, or
otherwise attempt to subvert the protocol.
Whereas, in a malicious failure model, faulty nodes may exhibit arbitrary,
potentially malicious, behavior.
We assume that a strong adversary can coordinate malicious nodes and
delay communication to compromise the replicated service.
However, the adversary cannot subvert standard cryptographic assumptions
about collision-resistant hashes, encryption, and signatures, e.g.,
the adversary cannot produce a valid signature of a non-faulty node.

Each pair of replicas is connected with point-to-point
bi-directional communication channels and each client can communicate
with any replica. Network links are pairwise authenticated, which guarantees
that a malicious replica cannot forge a message from a correct replica,
i.e., if replica $i$ receives a message $\mu$ in the incoming link from replica $j$,
then replica $j$ sent message $\mu$ to $i$ beforehand.

We use the state machine replication algorithm \cite{schneider1990implementing}
where replicas agree on
an ordering of incoming requests and all replicas execute the
requests in the same order.
Our system ensures safety in an asynchronous network that can
drop, delay, corrupt, duplicate, or reorder messages.
Liveness is guaranteed only during periods of synchrony when there is
a finite but possibly unknown bound on message delivery time.
The model puts no restrictions on clients, except that
their numbers must be finite,
however, safety and liveness require some constraints
on the number of faulty servers.

Depending on the role of a node and the type of message it wants to send,
messages may contain public-key signatures and message digests \cite{castro1999practical}.
A {\em message digest} is a numeric representation of the contents of a message
produced by collision-resistant hash functions.
Message digests are used to protect the integrity of 
a message and detect changes and alterations to any part of the message.
We denote a message $\mu$ signed by replica $r$ as
$\big\langle \mu \big\rangle_{\sigma_r}$ and
the digest of a message $\mu$ by $D(\mu)$.
For signature verification, we assume that
all machines have the public keys of all other machines.
In Section~\ref{sec:alg}, we explain when signatures and digests are needed.

\subsection{Quorum and Network Size}
We consider a cloud environment 
consisting of private and public clouds.
The system is an asynchronous distributed system containing
a set of $N$ replicas where
$S$ of them exist in a private cloud and
$P$ of them are in a public cloud.
Nodes in the private cloud are non-malicious:
either non-faulty (correct) nodes or crashed nodes.
The bound on the maximum number of crashed nodes is assumed to be $c$.
Similarly,
nodes in the public cloud can be either
non-faulty nodes or Byzantine nodes.
The bound on the maximum number of Byzantine nodes is $m$.
We call the nodes in the private cloud {\em trusted} and
the nodes in the public cloud {\em untrusted}.
All the clients and the replicas know which replicas
are trusted and which are untrusted.

Failures are divided into two disjoint classes:
malicious and crash faults.
In crash fault-tolerant models, e.g., Paxos \cite{lamport2001paxos}, 
given that $c$ nodes can crash, a request is
replicated to a quorum consisting of at least $c+1$ nodes to provide fault tolerance and 
to guarantee that a value once decided will remain decided in spite of failures (safety).
Furthermore, any two quorums intersect on at least \textit{one} node and
as a result, 
$2c +1$ is
the minimum number of nodes that allows an asynchronous system to
provide the safety and liveness properties.

In the Byzantine failure models, e.g., PBFT \cite{castro1999practical}, given that $m$ nodes can be
malicious, the quorum size should be at least $2m+1$ to ensure that
non-faulty replicas outnumber the malicious ones, i.e.,
a request is replicated in enough non-faulty nodes to guarantee safety in the presence of $m$ failures.
This implies that any two quorums intersect with at least $m+1$ nodes to ensure one correct node
in the intersection, thus the minimum network size is $3m+1$ \cite{bracha1985asynchronous}.

Likewise, in the hybrid model
the quorum size must include at least $2m+c+1$ nodes to tolerate $c$ crash and $m$ malicious failures~\cite{clement2009upright}.
This also guarantees that the intersection of any two quorums has to be at least $m +1$ nodes. 
Since the quorum size is $2m+c+1$ and the intersection of any two quorum $Q$ and $Q'$ is $m +1$ nodes,
${\mid} Q {\mid} + {\mid} Q' {\mid} = N + m +1 = 4m+2c+2$, thus
the (minimum) network size, $N$, will be~\cite{clement2009upright}
\begin{equation} \label{eq4}
N = 3m + 2c +1
\end{equation}

Intuitively, if there are $f$ failures
(of any type) in a network,
the network size has to be at least $f$ larger
than the quorum size, as any network with smaller size could lead to a deadlock
situation where none of the $f$ faulty servers are participating. 
Since, $f=m+c$ and the quorum size $Q$ is $2m+c+1$,
the network size should be at least $Q+f$ i.e., $3m+2c+1$.

\section{Public Cloud}\label{sec:public}

The hybrid failure model presented in Section~\ref{sec:model}
can be used by enterprises that own private clouds with a limited number of trusted servers
which is {\em insufficient} to run a fault-tolerant protocol.
This model gives them the option of renting from untrusted public clouds.
In this section, we present two methods to identify the number of servers
an enterprise needs to rent from a public cloud.

A business that owns
an insufficient number of trusted servers (servers that might crash but are not malicious) 
needs to rent more servers from some untrusted public clouds to satisfy
the minimum network size constraints ($3m +2c + 1$) of the protocol.
Public clouds might provide some statistics that show the percentage of faulty nodes in the cloud.
If there is no information on the type of failures, i.e. crash or malicious, within the public cloud,
we consider all the faulty nodes as malicious and
assume that the ratio of malicious nodes in public cloud ($m$) to
the size of public cloud ($\caP$) is known and is equal to $\alpha = \frac{m}{\caP}$.
Note that, we assume a uniform distribution of malicious nodes in public cloud, i.e.,
in any set $\pi \subseteq \caP$, at most $\alpha \times \pi$ nodes are malicious.

Given the size of the private cloud $S$, the bound on the maximum number of crashed nodes $c$ in the private cloud,
and the ratio $\alpha$ of malicious nodes ($m$) in the public cloud to
the size of the public cloud ($\caP$),
the task is to identify the required number of nodes $P$ to be rented from the public cloud that allows satisfying the protocol constraints.

The total number of nodes in the network is $N=S+P$.
Given our assumption of $\alpha$, we get $m = \alpha P$. Replacing $m$
in Equation~\ref{eq4}, we get
$N = 3 \alpha P + 2 c + 1$, which means,
$(3\alpha - 1) P = S - (2c+1)$, thus:
\begin{equation}\label{equ:p}
P = \lceil \frac{S - (2 c +1)}{3 \alpha -1} \rceil
\end{equation}
As an example consider the situation that a private cloud has $2$ servers
where one of them might be faulty, i.e., $S = 2$, and $c = 1$, and
we want to rent servers from a public cloud with $\alpha  = 0.3$.
Here, $P = \frac{2 - 2 -1}{3 * 0.3 -1} = \frac{-1}{-0.1} = 10$,
which means we need to rent $10$ servers from the public cloud to provide the safety constraints of the replication protocol.


In Equation~\ref{equ:p},
if the size of the private cloud ($S$) is equal or greater than $2c+1$,
then the private cloud does not need to rent any nodes and
can run a crash-fault tolerant protocol like Paxos \cite{lamport2001paxos} by itself.
If there is no private cloud ($S=0$) or 
all the nodes in the private cloud are faulty ($S=c$), using the private cloud has no advantage and
it is more reasonable to
rent all the required nodes from the public cloud and run a Byzantine fault-tolerant protocol in the public cloud. 
However, if $c < S < 2c+1$, renting some nodes from a public cloud might be helpful.

Similarly, if $\alpha \geq \frac{1}{3}$,
(i.e., more than $\frac{1}{3}$ of the nodes in the public cloud are malicious),
then the public cloud cannot satisfy the network size constraint
for Byzantine fault-tolerance.
Hence, an enterprise will need to rent servers if its private cloud size,
$S$, is between $c+1$ and $2c$, and
it can rent servers from public cloud providers that satisfy $\alpha < \frac{1}{3}$.
It should be noted that even if the size of the private cloud is equal or greater than $2c+1$, and
the public cloud does not satisfy the $\alpha < \frac{1}{3}$ constraint,
an enterprise might still rent some replicas from the public cloud for load balancing purposes.

Note that Equation~\ref{equ:p} can easily be extended to address the situation where 
the public cloud provides information on the ratio of both malicious and crash nodes, i.e.,
the ratio of malicious nodes to the size of public cloud ($\alpha = \frac{m}{\caP}$) as well as
the ratio of crash nodes to the size of public cloud ($\beta = \frac{c}{\caP}$) are known.
In such a situation, Equation~\ref{equ:p} can be rewritten as:
\begin{equation}\label{equ:p1}
P = \lceil \frac{S - (2 c +1)}{3 \alpha + 2 \beta -1} \rceil
\end{equation}

This method, which identifies the required number of nodes from a public cloud,
assumes a uniform distribution of faulty nodes in the public cloud.
However, public clouds might not guarantee a uniform distribution of $\alpha$ and alternatively
specify the maximum number of concurrent failures
in a cluster of rental nodes explicitly.
In such a setting, even if an enterprise rents a portion of that cluster, there is no guarantee that
the percentage of the faulty nodes in that portion is equal to the percentage of the faulty nodes in the entire cluster.
For example, a public cloud might guarantee that in a cluster of $10$ nodes, at most two concurrent failures can occur.
Nonetheless, if an enterprise rents only two nodes from that cluster, both of them might fail at the same time.
Assuming that the number of concurrent malicious failures in a
(cluster of nodes in a) public cloud is given and equal to $M$,
we would want to identify the required number of nodes $P$ to rent from such a public cloud.
The total number of nodes in the network is $N = 3m + 2c + 1 = S + P$ and
there is no guarantee on a uniform distribution of malicious nodes in the public cloud, thus $m = M$.
Hence, the required number of nodes is
$P = (3M + 2c + 1) - S$.

Similar to the first method, if the public cloud distinguishes between different types of failures and
provides information on the number of both crash and malicious failures, given as $C$ and $M$,
the required number of nodes from the public cloud is
$P = (3M + 2C + 2c + 1) - S$ where $c$, similar as before, is the number of crash failures in the private cloud.

Finally, it should be noted that both methods of identifying the public cloud size
can be generalized to multiple public clouds as well.
In Such a settings, since different public clouds might have different ratio (number) of faulty nodes,
the equation might have multiple solutions.
\section{\protocol}\label{sec:alg}

In this section we present \protocol,
a hybrid fault-tolerant consensus protocol
for a public/private cloud environment
that tolerates $m$ Byzantine failures in the public
and $c$ crash failures in the private cloud.

\protocol is inspired by the known Byzantine fault-tolerant protocol PBFT \cite{kapitza2012cheapbft}.
In PBFT, as can be seen in Figure~\ref{fig:pbft},
during a normal case execution, a client sends a request to a (primary) replica, and
the primary broadcasts a {\sf pre-prepare} message to all replicas.
Once a replica receives a valid {\sf pre-prepare} message, it broadcasts a {\sf prepare}
message to all other replicas.
Upon collecting $2f$ valid matching {\sf prepare} messages (including its own message)
that are also matched to the {\sf pre-prepare} message sent by the primary,
each replica broadcasts a {\sf commit} message.
In this stage, each replica knows that all non-faulty replicas
agree on the contents of the message sent by the primary.
Once a replica receives $2f+1$ valid matching {\sf commit} messages (including its own message),
it executes the request and sends the response back to the client.
Finally, the client waits for $f+1$ valid matching responses from different replicas
to make sure at least one correct replica executed its request.
PBFT also has a view change routine that provides liveness by allowing
the system to make progress when the primary fails.

\protocol consists of {\em agreement} and {\em view change} routines where
the agreement routine orders requests for execution by the replicas, and 
the view change routine coordinates the election of a new primary when
the current primary is faulty.

The algorithm, similar to most fault-tolerant algorithms,
is a form of state machine replication.
In such approaches, a service is replicated across a group of servers
in a distributed system.
Each server maintains a set of state variables, which are
modified by a set of "atomic" and "deterministic" operations.
Operations are {\em atomic} if they do not interfere with each other and
{\em deterministic} if the same operation executed in the same initial state
generates the same final state.
Also, the initial state must be the same in all replicas. 

The algorithm has to satisfy two main properties,
(1) {\em safety}: all correct servers execute the same requests in the same order, and
(2) {\em liveness}: all correct client requests are eventually executed.
Fischer et al. \cite{fischer1985impossibility} show that 
in an asynchronous system, where nodes can fail,
consensus has no solution that is both safe and live.
Based on that impossibility result, \protocol, similar to most fault-tolerant protocols,
ensures the safety property without any synchrony assumption and
considers a synchrony assumption to satisfy the liveness property.
Indeed, as long as the number of faulty nodes does not exceed the defined threshold,
a protocol can produce linearizable executions,
independent of whether the network loses, reorders, or arbitrarily delays messages.
However, a weak synchrony assumption is needed to satisfy liveness:
the delay from the moment when a request is sent by a client
for the first time and the moment when it is received by its
destination is in some fixed (but potentially unknown) interval.

We identify each replica using an integer in $[0,..., N{-}1]$ where
replicas in the private cloud, i.e., trusted replicas,
have identifiers in $[0, ..., S{-}1]$ and
replicas in the public cloud, i.e., untrusted replicas,
are identified using integers in $[S, ...,\allowbreak N{-}1]$.

In \protocol, the replicas move through a succession of configurations
called {\em views} \cite{el1985efficient}\cite{el1985availability}.
In a view, one replica is {\em the primary} and the others are {\em backups}.
Depending on the mode, some backups are {\em passive} and do not participate
in the agreement.
Views are numbered consecutively.
All replicas are initially in view $0$ and
are aware of their current view number at all time.

We explain \protocol in three different modes:
{\em Lion}, {\em Dog}, and {\em Peacock}
\footnote{ We call the modes Lion, Dog, and Peacock because seemorq (=SeeMoRe) is composed of these three animals.}.
In the Lion mode, the primary is {\em always} in the private cloud,
thus the primary is non-malicious.
The Dog mode is used to reduce the load on the private cloud
by assuming that the primary is still in the private cloud, but
instead of processing the client requests itself,
depends on $3m+1$ nodes in the public cloud to process the request.
This mode 
reduces the load on the private cloud, because except for the primary,
which does a single broadcast of the client's request,
other replicas in the private cloud are passive and do not participate in any phases.
Finally, in the Peacock mode, an untrusted node is chosen as the primary and
the protocol relies completely on the public cloud to process requests.
This mode is useful when
we intentionally rely completely on the public cloud for two purposes:
(1) load balancing when all the nodes in the private cloud are heavily loaded, or
(2) reducing the delay when there is a large network distance between the private and the public cloud and
the latency of having one more phase of communication within the public cloud
is less than the latency of exchanging messages between the two clouds.
The agreement routine of the Peacock mode is the same as PBFT \cite{kapitza2012cheapbft},
however, the view change routine can be more efficient.

In this section, we describe each of these three modes in detail,
followed by a technique to dynamically switch between the modes.
For each mode,
we first present the normal-case operation of the protocol and
then show how view changes are carried out
when it appears that the primary has failed.
Next, we show how \protocol can
dynamically switch between these three modes.
We use $\pi$ to show the current mode of the protocol where $\pi \in \{1,2,3\}$.
At the end of this section, we also present a short discussion on different modes of \protocol and compare it
with some known relevant protocols, i.e.,
the crash fault-tolerant protocol Paxos \cite{lamport2001paxos},
the Byzantine fault-tolerant protocol PBFT \cite{castro1999practical}, and
the hybrid fault-tolerant protocol UpRight \cite{clement2009upright}.

\subsection{The Lion Mode: Trusted Primary} \label{sec:mode1}

Owning a private cloud gives \protocol the chance to choose a trusted node
as the primary.
When the primary is trusted, all the non-faulty backups receive correct
messages from the primary, which
eliminates the need to multicast messages by replicas to
realize whether all the non-faulty ones receive the same message or not.
Thus, we can reduce one phase of communication and
a large number of messages.

In particular, within a view, the normal case operation for \protocol
to execute a client request in the Lion mode proceeds as follows.
A client sends a \req message to the primary,
i.e., a trusted replica in the private cloud.
The primary assigns a sequence number to the request and
multicasts a \one message including the request to all replicas.
Replicas receive a \one from the primary and send an \two to the primary.
The primary upon receiving $2m+c+1$ matching \two messages,
sends a \three message to all replicas and a \reply to the client.
Upon receiving a \three message from the primary, replicas execute the client request.
Finally, the client receives a \reply message from the primary and marks the request as complete.

\begin{figure*}[t]
  \centering
  \subfigure[The Lion Mode]{
  \label{fig:case1}
  \includegraphics[width=0.22\linewidth]{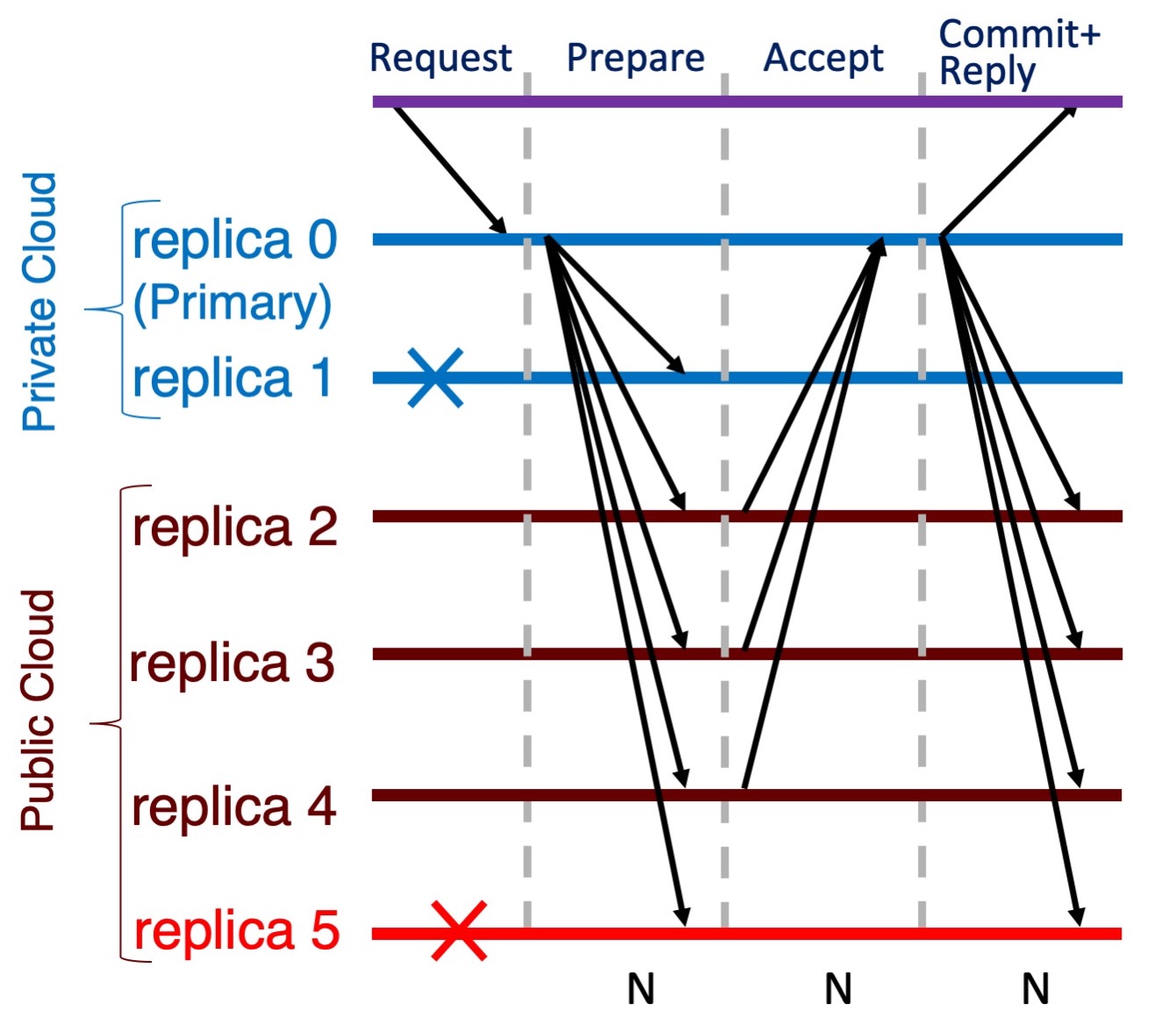}}
  \subfigure[The Dog Mode]{
  \label{fig:case2}
  \includegraphics[width=0.22\linewidth]{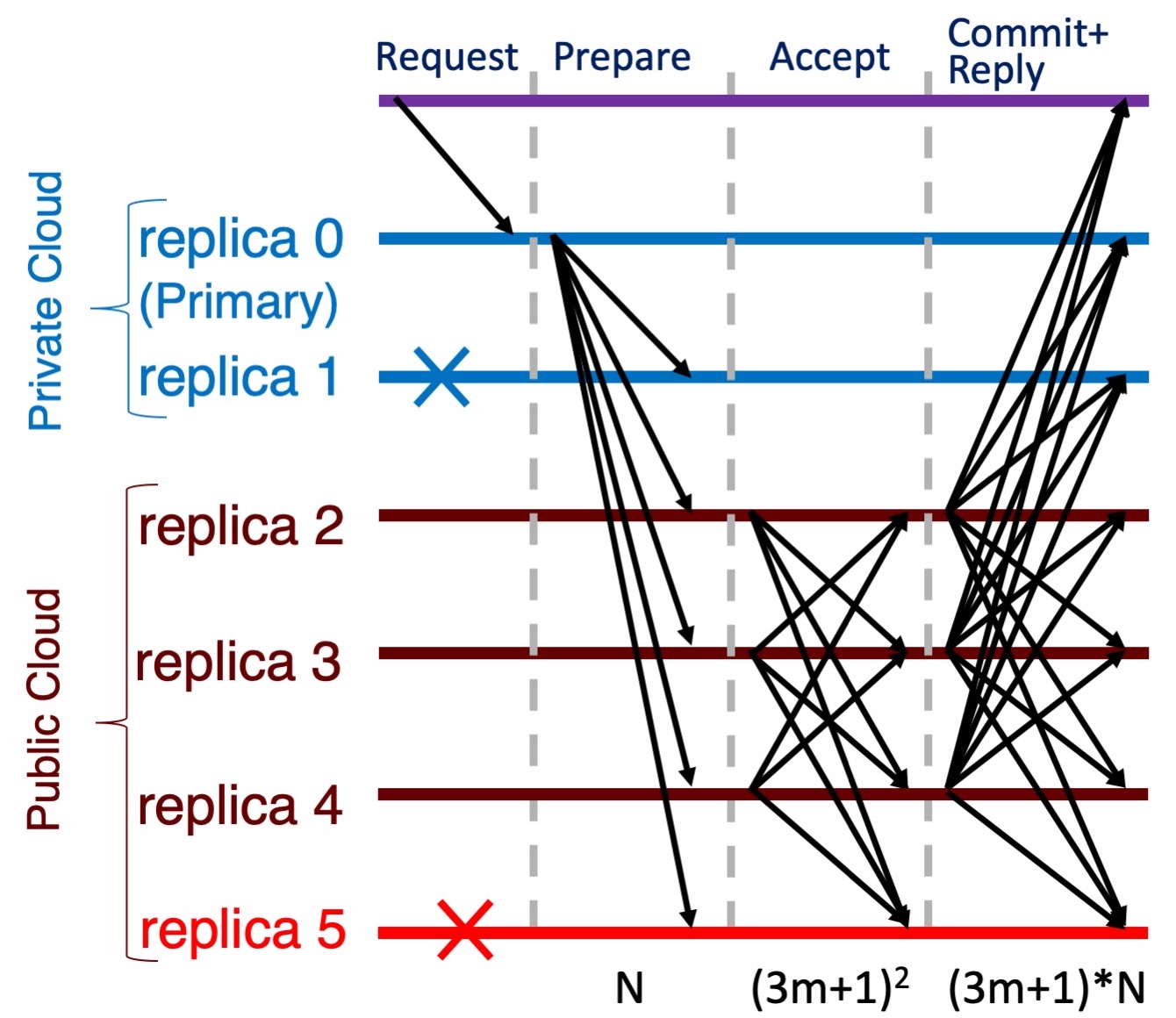}}
  \subfigure[The Peacock Mode]{
  \label{fig:case3}
  \includegraphics[width=0.25\linewidth]{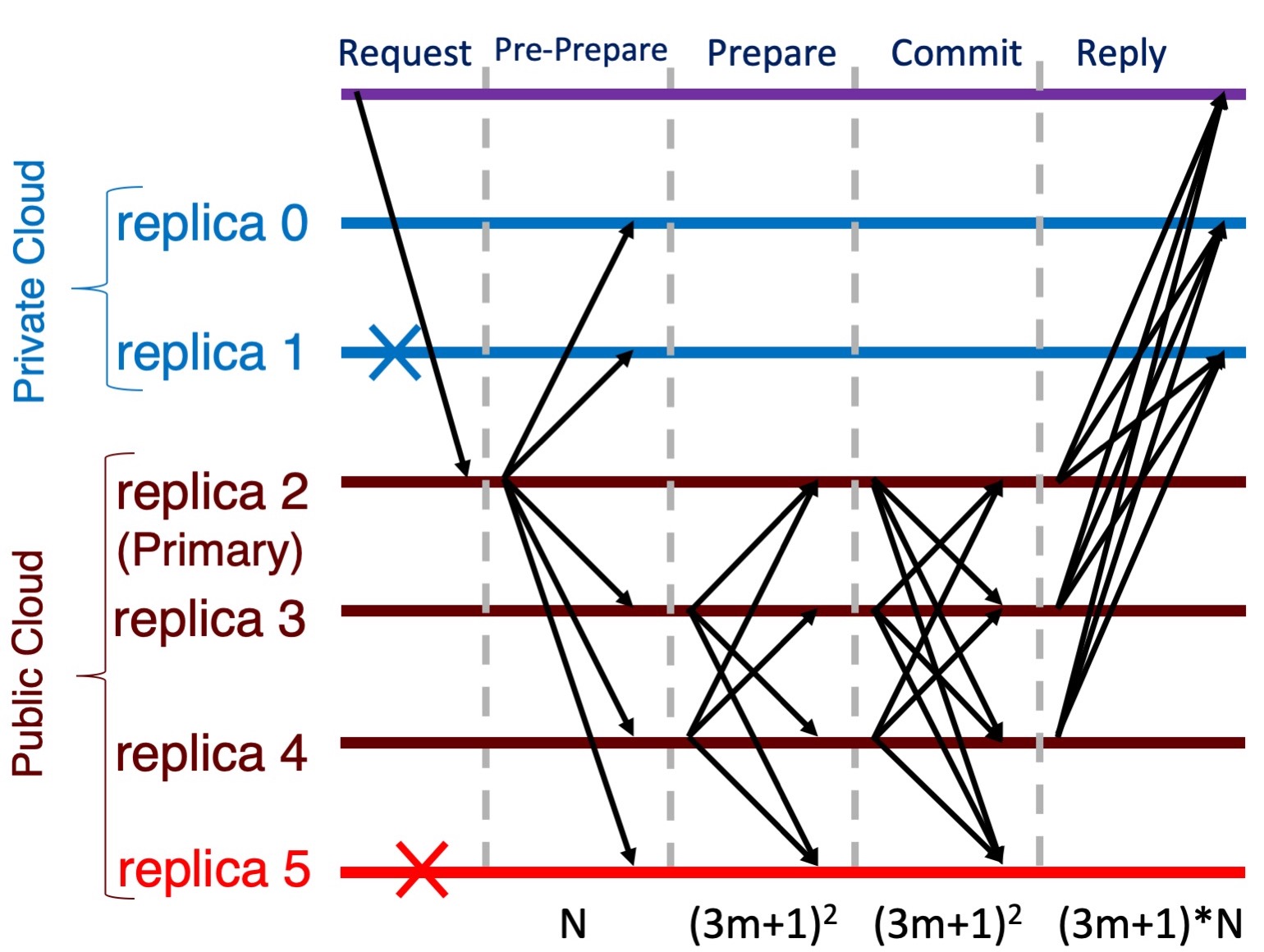}}
  \subfigure[PBFT]{
  \label{fig:pbft}
  \includegraphics[width=0.23\linewidth]{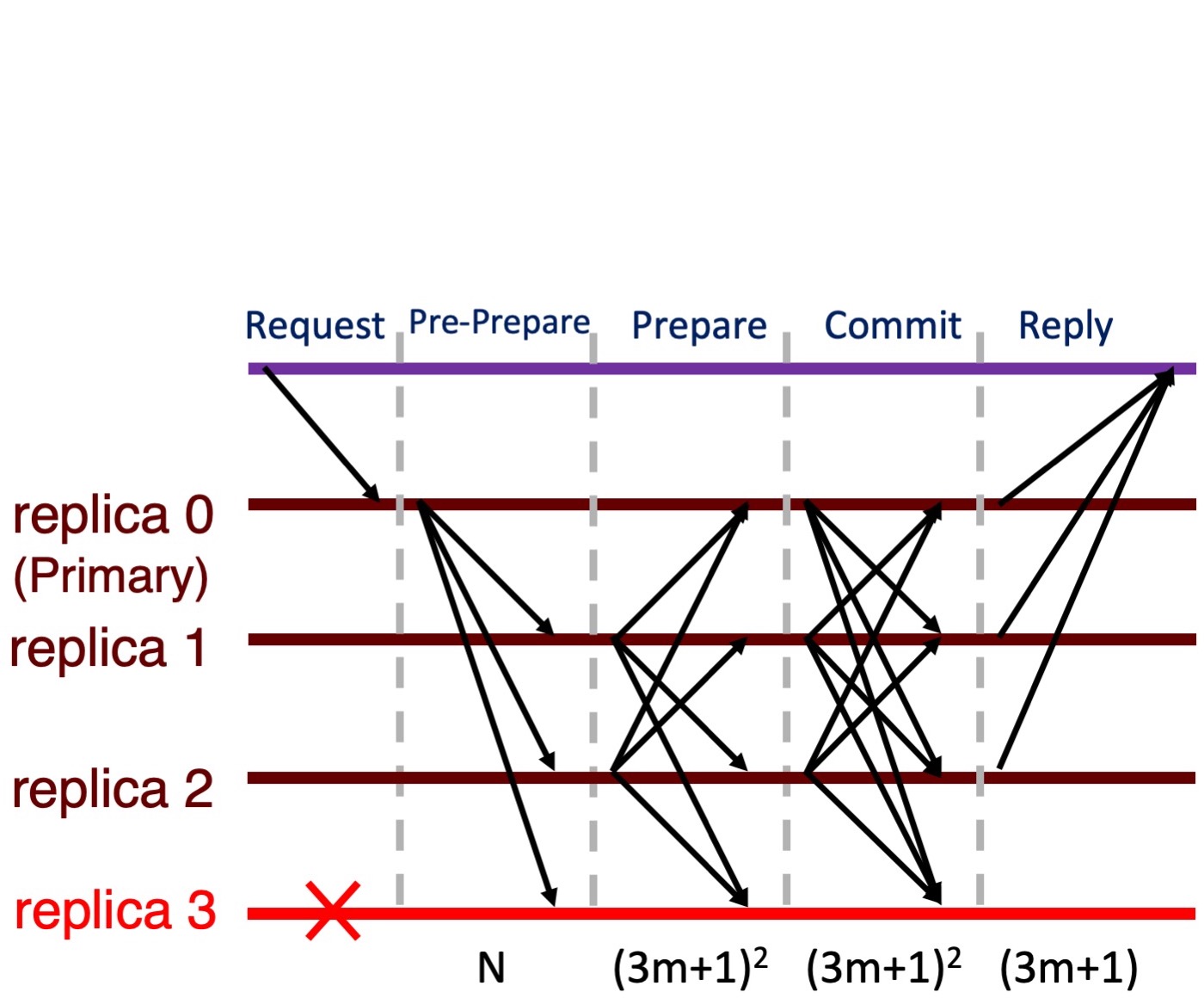}}
  \caption{The normal case operation of the three modes of \protocol and PBFT}
  \label{fig:normalcase}
  \end{figure*}

Figure~\ref{fig:case1} shows the normal case operation of the Lion mode.
Here, replicas $0$ and $1$ are trusted ($S=2$) and
the four other replicas, $2$ to $5$, are untrusted ($P=4$).
In addition, one of the trusted replicas $(1)$ is crashed ($c=1$) and
one of the untrusted replicas $(5)$ is malicious ($m=1$).
With a trusted primary, the total number of exchanged messages is $3N$.

The pseudo-code for the Lion mode is presented in Algorithm~\ref{alg:lion}.
Although not explicitly mentioned, every sent and received message is logged
by the replicas.
Each replica is initialized with a set of variables as indicated in lines
1-4 of the algorithm. The primary of view $v$ is a replica $p$ such that
$p = (v \mod S)$. A client $\varsigma$ requests a state machine operation $op$
by sending a message $\langle\text{\scriptsize \REQ}, op, ts_{\varsigma}, \varsigma \rangle_{\sigma_{\varsigma}}$
to replica $p$ it believes to be the primary.
The client's timestamp $ts_\varsigma$ is used to totally order the requests and to
ensure exactly-once semantics. The client also signs the message with signature
$\sigma_{\varsigma}$ for authentication.

Each replica keeps the state of the service,
a message log containing valid messages the replica has received, and
two integers denoting the replica's current view and mode numbers.
Message logs then serve as the basis for maintaining consistency in view changes.

As indicated in lines 5-8, upon receiving a client \req,
the primary $p$ first checks if the signature and timestamp in the
request are valid and simply discards the message otherwise.
The primary assigns a sequence number $n$ to the request and multicasts a signed
$\langle\langle\text{\scriptsize \ONE}, v, n, d \rangle_{\sigma_p}, \mu \rangle$
message to all the replicas
where $v$ is the current view, $\mu$ is the client's request message, and
$d$ is the digest of $\mu$.
At the same time, the primary appends the message to its log.
The primary signs its message, because it might be used by other replicas
later in view changes as a proof of receiving the message.

As shown in lines 9-11 of the algorithm, upon receipt of
$\langle\langle\text{\scriptsize \ONE}, v, n, d \rangle_{\sigma_p}, \mu \rangle$
from primary $p$, replica $r$ checks
if view $v$ is equal to the replica's view. It then logs the \one message,
and responds to the primary with
$\langle\text{\scriptsize \TWO}, v, n, d, r \rangle$ message.
Since \two messages are sent only to the trusted primary
and are not used later for any other purposes,
there is no need to sign these messages.

Upon collecting $2m+c$ valid \two messages from different replicas
(plus itself becomes $2m+c+1$)
for the request $\mu$ in view $v$ with sequence number $n$, as seen in
lines 12-15,
the primary multicasts a \three message
$\langle \langle\text{\scriptsize \THREE}, v, n, d \rangle_{\sigma_p}, \mu \rangle$
to all replicas.
The primary attaches the request $\mu$ to its \three message,
so that if a replica has not received a \one message for that request,
it can still execute the request.
The primary also executes the operation $op$ and sends a \reply message
$\langle\text{\scriptsize \REPLY}, \pi, v, ts_{\varsigma}, u  \rangle_{\sigma_p}$
to client $\varsigma$.
Mode number $\pi$ and view number $v$ are sent to clients to 
enable them to track the current mode and view and hence the current primary.
It is important especially when a mode change or view change occurs, replacing the
primary.

\begin{algorithm}[t]
\scriptsize
\caption{The Normal-Case Operation in the {\em Lion} mode}
\label{alg:lion}
\begin{algorithmic}[1]
\State \textbf{init():} 
\State    \quad $r$ := {\sf replicaId}
\State    \quad $v$ := {\sf viewNumber}
\State    \quad if $r = (v$ mod $S)$ then {\sf isPrimary} := \textsf{true} \newline

\State \textbf{upon receiving} $\mu {=} \langle\text{\REQ}, op, ts_{\varsigma}, \varsigma \rangle_{\sigma_{\varsigma}}$
and {\sf isPrimary}:

\State    \quad if $\mu$ is {\em valid} then
    
\State    \qquad {\bf assign} {\em sequence number} $n$
    
\State    \qquad \textbf{send} $\langle\langle\text{\ONE}, v, n, d \rangle_{\sigma_p}, \mu \rangle$
to all {replicas}\newline

\State \textbf{upon receiving} $\langle\langle\text{\ONE}, v, n, d \rangle_{\sigma_p}, \mu \rangle$ from primary $p$:

\State    \quad if $v$ is {\em valid} then
    
    
\State    \qquad \textbf{send} $\langle\text{\TWO}, v, n, d, r \rangle$ to {primary} $p$\newline

\State \textbf{upon receiving} $\langle\text{\TWO}, v, n, d, r \rangle$ from \textbf{2m{+}c} replicas and {\sf isPrimary}:

\State    \quad \textbf{send} $\langle \langle\text{\THREE}, v, n, d \rangle_{\sigma_p}, \mu \rangle$ to all replicas

\State \quad {\bf execute} operation $op$
\State    \quad \textbf{send} $\langle\text{\REPLY}, \pi, v, ts_{\varsigma}, u  \rangle_{\sigma_p}$
to client $\varsigma$ with result $u$
\end{algorithmic}
\end{algorithm}

Once a replica receives a valid \three message with correct
view number from the primary, it executes the operation $op$, if all 
requests with lower sequence numbers than $n$ has been executed.
This ensures that all non-malicious replicas execute requests
in the same order as required to provide the safety property.
Note that even if the replica has not received a \one message for that request,
as long as the view number is valid and
the message comes from the primary,
the replica considers the request as committed.

When the client receives a \reply message
$\langle\text{\scriptsize \REPLY}, \allowbreak \pi, v, ts_{\varsigma}, u  \rangle_{\sigma_p}$
with a valid signature
from primary $p$
and with the same timestamp as the client's request,
it accepts $u$ as the result of the requested operation.

If the client does not receive a \reply from the primary after a preset time, the client
may suspect a crashed primary. The client then broadcasts the same request to all replicas.
A replica, upon receiving the client's request, checks if it has already executed the request;
if so, it simply sends the \reply message to the client.
The client waits for a \reply from the private cloud or
$m+1$ matching \reply messages from the public cloud before accepting the result.
The primary will eventually be suspected to be faulty by enough replicas to trigger a view change.

\medskip
\noindent{\bf State Transfer}.
A fault-tolerant protocol must provide a way to checkpoint the state of different replicas.
It is especially required in an asynchronous system where even non-faulty replicas
can fall arbitrarily behind.
Checkpointing also brings slow replicas up to date
so that they may execute more recent requests.
Similar to \cite{castro1999practical}, in our protocol,
checkpoints are generated periodically when a request
sequence number is divisible by some constant
(checkpoint period).

Trusted primary $p$ produces a checkpoint and multicasts a
$\langle\text{\scriptsize \CHECKP}, n,d \rangle_{\sigma_p}$ message
to the other replicas, where
$n$ is the sequence number of the
last executed request and $d$ is the digest of the state.
A server considers a checkpoint to be {\em stable} when it
receives a \checkp message for sequence number $n$
signed by trusted primary $p$.
We call this message a {\em checkpoint certificate}, which proves that
the replica's state was correct until that request execution.

Checkpointing not only brings slow replicas up to date,
but it can also be used as a garbage collection mechanism.
All the messages sent by a replica are kept in a message log in case
they have to be re-sent.
However, when a checkpoint becomes stable,
replicas do not need to keep messages prior to
the checkpoint in their log and can simply discard
all \one, \two, and \three messages with sequence numbers
less than or equal to the checkpoint's sequence number.
They also discard all earlier checkpoints and checkpoint messages.

\medskip
\noindent{\bf View Changes}.
The goal of the view change protocol is to provide liveness by allowing
the system to make progress when a primary fails.
It prevents replicas from waiting indefinitely for requests to execute.
A view change must guarantee that it will not introduce any changes in a
history that has been already completed at a correct client.
Most view change routines
\cite{el1985availability, el1985efficient, castro1999practical, kotla2007zyzzyva, yin2003separating, castro2003base, cowling2006hq}
are triggered by timeouts and
require enough non-faulty replicas to exchange view change messages.
\protocol uses a similar technique in the Lion mode.
In such a situation, replicas detect the failure and
reach agreement to change the view from $v$ to $v'$.
The primary of new view $v'$ then handles the 
uncommitted requests,
and takes care of the new client requests.

View changes are triggered by timeout.
When a replica receives a valid \one message from the primary,
it starts a timer that expires after some defined time $\tau$.
When the backup receives a valid \three message, the timer is stopped,
but if at that point the backup is waiting for a \three message for
some other request, it restarts the timer.
If the timer expires, the backup suspects that the primary is faulty and
starts a view change.

When a backup suspects that the primary is faulty
(its timer for some \one message expires),
it stops accepting \one and \three messages and multicasts a
$\langle\text{\scriptsize \VCHANGE}, \allowbreak v+1, n, \xi, \cal P, \cal C \rangle$
message to all replicas where
$n$ is the sequence number of the last stable checkpoint known to $r$,
$\xi$ is the checkpoint certificate, and
$\cal P$ and $\cal C$ are two sets of received valid \one 
(without the request message $\mu$)
and \three messages
for requests with a sequence number higher than $n$.

When primary $p'$ of new view $v+1$ receives
$2m+c$ valid \vchange messages
from different replicas, it multicasts a
$\langle\text{\scriptsize \NEWV}, v+1, \cal P', \cal C'$ $\rangle_{\sigma_{p'}}$
message to all replicas where
$\cal P'$ and $\cal C'$ are two sets of
\one and \three messages respectively
which are constructed as follows.
 
Let $l$ be the sequence number of the latest checkpoint, and
$h$ be the highest sequence number of a \one message in all the received $\cal P$ sets.
For each sequence number $n$ where $l < n \leq h$,
the primary does the following steps:

\begin{enumerate}
\item It first checks all \three messages in set $\cal C$ of the received \vchange messages.
If the primary finds a \three message with a valid signature $\sigma_p$
($p$ was the primary of view $v$)
for some request $\mu$
the primary adds a
$\langle\langle\text{\scriptsize \THREE}, v+1, n, d  \rangle_{\sigma_{p'}}, \mu\rangle$
to $\cal C'$

\item If no such \three message is found,
the primary checks the \one messages in $\cal P$ sets:

$\bullet$ If the primary finds $2m+c+1$ valid \one messages for $n$,
it adds a
$\langle\langle\text{\scriptsize \THREE}, v+1, n, d \rangle_{\sigma_{p'}}, \mu\rangle$
to $\cal C'$.

$\bullet$ Else, if it receives at least one valid \one message for $n$,
the primary adds a
$\langle \langle\text{\scriptsize \ONE}, v{+}1, n, d \rangle_{\sigma_{p'}},\mu\rangle$
to $\cal P'$.

\item If none of the above situations occur,
there is no valid request for $n$, so
the primary adds a
$\langle\text{\scriptsize \ONE}, v{+}1,\allowbreak n, d \rangle_{\sigma_{p'}},
\mu^{\emptyset}\rangle$
to $\cal P'$ where $\mu^{\emptyset}$ is a special {\em no-op} command
that is transmitted by the protocol like other requests but leaves the state unchanged.
The third situation happens when no replica has received
a \one message from the previous primary.

\end{enumerate}

In contrast to PBFT, 
since the primary is trusted,
it does not need to append
all the \vchange messages in the \newv message which makes the \newv messages
much smaller.
The primary inserts all the messages in $\cal P'$ and $\cal C'$
to its log.
It also checks the log to make sure its log contains the latest stable checkpoint.
If not, the primary inserts \checkp messages for the checkpoint $l$ and
discards the earlier information from the log.

Once a replica in view $v$ receives a \newv message from the primary of view $v+1$,
the replica logs all \one and \three messages,
updates its checkpoint in the same way as the primary, and
for each \one message, sends an \two message to the primary.
Non-faulty replicas in view $v$ will not accept a \one message for a new view $v' > v$
without having received a \newv message for $v'$.

\medskip
\noindent{\bf Correctness}.
Within a view, since the primary is trusted and it assigns sequence numbers to the requests,
safety is ensured as long as the primary does not fail.
Indeed, for any  two committed requests $r_1$ and $r_2$ with sequence numbers $n_1$ and $n_2$ respectively,
if $D(r_1) = D(r_2)$, then $n = n'$.

If the primary fails a view change is executed.
To ensure safety across views,
the primary waits for $2m+c$ \two messages (considering itself, a quorum of $2m+c+1$)
from different replicas to ensure that committed requests are totally ordered across views.
In fact, for any  two committed requests $r_1$ and $r_2$ with sequence numbers $n_1$ and $n_2$,
since a quorum of $2m+c+1$ replicas commits $r_1$ and a quorum of $2m+c+1$ replicas commits $r_2$, and
these two quorums have at least $m+1$ overlapping nodes,
there should be at least one non-faulty node that commits both $r_1$ and $r_2$
but this is not possible because the node is not faulty.
As a result, if $D(r_1) = D(r_2)$, then $n = n'$.
This guarantees that in the event of primary failure,
any new quorum of $2m+c+1$ replicas will have at least $m+1$ overlapping nodes
that received a \one message (and sent \two) for request $\mu$ from the previous primary.
Thus, there is at least one non-faulty node in that quorum that helps
the protocol to process request $\mu$ in the new view.

\subsection{The Dog Mode: Trusted Primary, Untrusted Backups}\label{sec:mode2}

The Dog mode is proposed to reduce the load on the private cloud.
In this mode,
a {\em trusted primary} receives a \req message, assigns a sequence number, and
relies on $3m+1$ {\em untrusted nodes} (in the public cloud) to process the request. These
$3m+1$ nodes are called \textit{proxies}.
Since a trusted primary assigns the sequence number to the request before broadcasting,
this reduces the scope of any malicious behaviour.
Whereas in PBFT, when replicas receive a message from the primary,
they perform one round of communication to make sure
all non-faulty replicas agree on a total order for the requests within a view.
However, here, since a trusted primary assigns the sequence numbers, similar to the Lion mode,
there is no need for that phase.

Figure~\ref{fig:case2} shows the normal case operation of \protocol
with a trusted primary (node $0$).
As before, two replicas are trusted ($S=2$), four replicas are untrusted ($P=4$), $c=1$, and $m=1$.
Since a trusted primary assigns sequence numbers, the protocol, similar to Paxos, needs two phases to process requests.
However, since the protocol tolerates malicious failures, the number of messages in terms of the number of replicas,
similar to PBFT, is quadratic.
Here, there are totally $N+(3m+1)^2+ (3m+1)*N$ messages exchanged where $3m+1$ is the total number of proxies.
In this particular example, since $m=1$, all replicas in the public cloud are proxies.

Algorithm~\ref{algo:dog} provides the pseudo-code for the Dog mode.
Lines 1-5 indicate the initialization of state variables for each replica,
including the primary and the proxies.
A replica $r$ in the public cloud is a {\em proxy} in view $v$ if
$r{-} (v \mod P) {\in} [S, ...,S{+}3m]$.
Here since replicas are in the public cloud, $r$ is an integer in $[S, ..., N{-}1]$.
The public cloud might have more than $3m{+}1$ replicas,
however, $3m{+}1$ is enough to reach consensus and
any additional replicas may degrade the performance.
The trusted primary of view $v$ is chosen in the same way as the first mode, i.e., $p$ is the primary if
$p {=} (v \mod S)$.

As shown in lines 6-9 of the algorithm, the primary, upon receiving
request $\mu$, validates the timestamp and signature of $\mu$,
assigns a sequence number $n$, and multicasts signed \one message
$\langle\langle\text{\scriptsize \ONE}, v, n, d \rangle_{\sigma_p}, \mu \rangle$
to all replicas.

When a proxy receives a \one message from the primary, as indicated in
lines 10-12,
it validates the view number,
logs the message and sends a signed \two message
$\langle\text{\scriptsize \TWO}, v, n, d, r \rangle_{\sigma_r}$
to all the other proxies.
Here, in contrast to the first mode,
the proxy signs its message as a proof of message reception in case
of a view change.

\begin{algorithm}[t]
\caption{The Normal-Case Operation in the {\em Dog} mode}
\label{algo:dog}

\begin{algorithmic}[1]
\scriptsize

\State \textbf{init():} 
\State    \quad $r$ := {\sf replicaId}
\State    \quad $v$ := {\sf viewNumber}
\State    \quad if $r = (v \mod S)$ then {\sf isPrimary} := \textsf{true}
\State    \quad else if $r - (v \mod P)$ $\in [S,..,S+3m]$ then {\sf isProxy} := \textsf{true}\newline

\State \textbf{upon receiving} $\mu = \langle\text{\REQ}, op, ts_{\varsigma}, \varsigma \rangle_{\sigma_{\varsigma}}$ and {\sf isPrimary}:

\State    \quad if $\mu$ is {\em valid} then
    
\State    \qquad {\bf assign} {\em sequence number} $n$
    
\State    \qquad \textbf{send} $\langle\langle\text{\ONE}, v, n, d \rangle_{\sigma_p}, \mu \rangle$ to all {replicas}
\newline

\State \textbf{upon receiving} $\langle\langle\text{\ONE}, v, n, d \rangle_{\sigma_p}, \mu \rangle$ from the primary $p$ and {\sf isProxy}:

\State    \quad if $v$ is {\em valid} then

\State    \qquad \textbf{send} $\langle\text{\TWO}, v, n, d, r \rangle_{\sigma_r}$ to all {proxies}
\newline

\State \textbf{upon receiving} $\langle\text{\TWO}, v, n, d, r \rangle$ from \textbf{2m+1} proxies:

\State    \quad \textbf{send} $\langle\text{\THREE}, v, n, d, r \rangle_{\sigma_{r}}$ to all other proxies
    
\State    \quad \textbf{send} $\langle\text{\INFORM}, v, n, d, r \rangle_{\sigma_r}$ to all private cloud nodes and non-proxy nodes in public cloud

\State \quad {\bf execute} operation $op$

\State    \quad \textbf{send} $\langle\text{\REPLY}, \pi, v, ts_{\varsigma}, u  \rangle_{\sigma_r}$
to client $\varsigma$ with result $u$

\end{algorithmic}
\end{algorithm}

As described in lines 13-17 of the algorithm, 
upon receiving $2m+1$ matching \two messages (including its own message)
with correct signatures,
a proxy $r$ multicasts a \three message
$\langle\text{\scriptsize \THREE}, v, n, d, r \rangle_{\sigma_{r}}$
to the other proxies.
Each proxy $r$ also sends a signed \inform message
$\langle\text{\scriptsize \INFORM}, v, n, r, d \rangle_{\sigma_r}$
to all the nodes in the private cloud and all non-proxy nodes in the public cloud.
The \inform message, including its
identifier $r$ and message digest $d$,
to inform them that such a request is committed.
Non-proxy nodes wait for $2m+1$ valid matching \inform messages from different proxies
which are matched by the \one message that they received from the primary
before executing the request. 
If the proxy has executed all requests with sequence numbers lower than $n$, it
executes the request $n$ and sends a \reply message
$\langle\text{\scriptsize \REPLY}, \pi, v,\allowbreak ts_{\varsigma}, u  \rangle_{\sigma_r}$
with result $u$ to the client.

Any other replica that receives $m+1$ matching \three messages
from the proxies with
valid signatures,
correct message digest, and
view numbers equal to its view number
considers the request as committed, and executes the request.
Since all the replicas receive \one messages from the primary,
they have access to the request and can execute it.

The client also waits for $2m+1$ matching \reply messages
from different proxies before accepting the result.
If the client has not received a valid reply after a preset time,
the client multicasts the request to the proxies.
The proxies re-send the result if the request has already
been processed and the client waits for $m+1$ matching \reply messages from the proxies
before accepting the result.
Otherwise, similar to the first mode,
eventually the primary will be suspected to be faulty by enough replicas
and a view change will be triggered.

\medskip
\noindent{\bf State Transfer}.
Checkpointing in the Dog mode works in the same way as
the Lion mode.
Trusted primary $p$ multicasts a signed \checkp message
to all other replicas with the sequence number of the
last executed request and the digest of the state.
Upon receiving a \checkp message from the primary,
a server considers that a checkpoint is {\em stable} and logs
the message which is used later as a
{\em checkpoint certificate}.

\medskip
\noindent{\bf View Changes}.
In the Dog mode, view change happens when the trusted primary is suspected to have crashed.
Here, similar to the Lion mode, we rely on the primary of new view to
handle the prepared but not yet committed requests.
However, since the nodes in the public cloud are processing
the requests, they are the ones who send \vchange messages.
Each node in the public cloud
multicasts a \vchange message
$\langle\text{\scriptsize \VCHANGE}, v+1, n, \xi, \cal P \rangle$
to all the nodes in the public cloud and the primary of the next view where
$\xi$ is the checkpoint certificate for sequence number $n$,
and
$\cal P$ is
the set of received valid \one messages
with a sequence number higher than $n$.

In this mode, in contrast to the Lion mode,
nodes do not include the set of \three messages ($\cal C$) in their \vchange messages because
in the Dog mode, to show that a request is committed, a nodes needs to include $2m+1$ valid \three messages for that request,
which makes the \vchange messages much larger.

Primary $p'$ of the new view waits for $2m+1$ valid \vchange messages
from the proxies of the last active view, i.e., the view with a non-faulty primary, before
multicasting a \newv message.
This is needed to ensure the correctness of the protocol
even if there are consecutive crashed primary nodes (inactive views) and
the number of nodes in the public cloud are more than the number of proxies
(the set of proxies are changed from one view to another).

Upon receiving $2m+1$ valid \vchange messages, primary $p'$ of view $v+1$ multicasts a \newv message
$\langle\text{\scriptsize \NEWV}, \allowbreak v+1, \cal P'$ $\rangle_{\sigma_{p'}}$
to all the replicas where
for each sequence number $n$
(between the latest checkpoint and the highest sequence number of a \one message),
if there is any valid \one message in set $\cal P$
of the received \vchange messages,
the primary adds a
$\langle\text{\scriptsize \ONE}, v+1, n, d  \rangle_{\sigma_{p'}}$
to $\cal P'$.
Else, there is no valid request for $n$, so
similar to the Lion mode, 
the primary adds a no-op \one message
$\langle\text{\scriptsize \ONE}, v+1, n, d \rangle_{\sigma_{p'}}, \mu^{\emptyset}\rangle$
to $\cal P'$.

Here, again, since the primary is trusted it does not need
to include \vchange messages in the \newv message.
The primary then inserts all the messages in $\cal P'$
to its log and updates its checkpoint, if needed.

Once a proxy of view $v+1$ receives a \newv message from the primary of view $v+1$,
the proxy logs all \one messages,
updates its checkpoint, and
multicasts an \two message to all the proxies
for each \one message in $\cal P'$.
Other replicas also receive the \newv message to be informed that the view is changed.

\medskip
\noindent{\bf Correctness}.
Within a view, since the primary is trusted and it assigns sequence number to the requests,
similar to the Lion mode, safety is ensured as long as the primary does not fail.
To ensure safety across views,
since $3m+1$ nodes participate in the protocol, to commit a message,
$2m+1$ matching \two messages are needed.
In fact, for any  two committed requests $r_1$ and $r_2$ with sequence numbers $n_1$ and $n_2$,
since a quorum of $3m+1$ replicas commits $r_1$ and a quorum of $3m+1$ replicas commits $r_2$, and
these two quorums have at least $m+1$ overlapping nodes,
there is at least one non-faulty node that commits both $r_1$ and $r_2$,
but this is not possible because the replica is non-faulty.
As a result, if $D(r_1) = D(r_2)$, then $n = n'$.

\subsection{The Peacock Mode: Untrusted Primary, Untrusted Backups}\label{sec:mode3}

One characteristic of online services is the ever changing patterns in client requests.
While there might be periods of high traffic thus overloading some servers, at other 
periods, the resources may be underutilized.
Also, depending on server placements and communication delays,
enterprises may benefit from protocols
that allow a subset of the servers, e.g. only the public cloud, to handle certain client requests.

The third mode of the protocol, the Peacock mode, is presented to handle two different situations.
First,  when the private cloud is heavily loaded and the public cloud can
handle the requests by itself for load balancing.
Second, when there is a large network distance between the private
and the public cloud and
the latency due to one more phase is less than the latency of
exchanging messages between the two clouds.
In both situations, the nodes in the private cloud become passive replicas in the agreement routine
and are only informed about the committed messages.
However, they still may participate in the view change routine.

In the Peacock mode, \protocol completely relies on $3m+1$ nodes in the public cloud to
process the requests.
The untrusted primary of view $v$ in the Peacock mode
is replica $p$ where $p = (v \mod P) + S$.
Similar to the Dog mode, since there might be more than $3m+1$ replicas in the public cloud,
in each view, $3m+1$ are chosen as proxies.
Node $i$ is a proxy in view $v$ if $i- (v \mod P) \in [S, ...,S+3m]$.
This ensures that the primary is always a proxy.

In the Peacock mode, \protocol processes the requests using PBFT \cite{castro1999practical}
with two small changes.
First, the primary multicasts signed \zero message along with the request to all the nodes
(and not only the $3+1$ proxies).
Second, when the request is committed, each proxy $r$ sends a signed \inform message
$\langle\text{\scriptsize \INFORM}, v, n, d, r \rangle_{\sigma_r}$
to all the nodes in the private cloud as well as all non-proxy nodes in the public cloud.
Other nodes also wait for $m+1$ valid matching \inform messages
from different proxies before executing the request.

As indicated in Figure~\ref{fig:case3},
similar to PBFT, the Peacock mode processes the requests in three phases:
{\sf \zero}, {\sf \one}, and {\sf \three}.
As can be seen, the replicas in the private cloud have no participation
in any phases and are only informed about the committed requests.
The total number of exchanged messages
in the Peacock mode is $N+ 2 * (3m+1)^2 + (1+S) * (3m+1)$.


\medskip
\noindent{\bf View Changes}.
In the Peacock mode, we rely on a trusted node in the private cloud, called {\em transferer}, to change the view.
Indeed, instead of the primary of the new view, a transferer changes the view.
Replica $t$ in the private cloud is the transferer of view $v'$
(changes the view from $v$ to $v'$)
if $t = (v'\mod S$).
Choosing a transferer to change views helps in
minimizing the size of \newv messages and
more importantly, reduces the delay between the request and its reply.
Because even if there are consecutive malicious primary nodes,
since the transferer takes care of the uncommitted requests of view $v$,
the protocol does not carry the messages from one view to another.
In contrast, in PBFT, it is possible that a valid request in view $v$
be committed in view $v+m$ (when there are $m$ consecutive primaries).
Other than the transferer, view change in the Peacock mode is similar to PBFT.
Proxies multicast \vchange messages to all replicas.
When the transferer of new view $v+1$ receives
$2m+1$ valid \vchange messages
from different proxies of view $v$,
it multicasts a \newv message to all replicas in both public and private clouds.
Once a proxy receives a valid \newv message, it logs all the \one messages,
updates its checkpoint, and sends an \two message to all other proxies
for each \one message.
When the transferer has changed the view and the new primary receives the \newv message from the transferer,
the new primary starts to process new requests in view $v+1$.

\medskip
\noindent{\bf Correctness}.
In the Peacock mode, the protocol ensures safety and liveness similar to PBFT \cite{castro1999practical}.

\subsection{Dynamic Mode Switching}\label{sec:switch}

We presented three different modes of \protocol and
explained when each mode is useful.
Now, we show how to dynamically switch between different modes.

An enterprise might prefer to use the Lion mode of \protocol,
because it needs fewer phases
(in comparison to the Peacock mode) and
less number of message exchanges
(in comparison to the Dog or Peacock mode).
However, if the private cloud becomes heavily loaded, or
at some point, a high percentage of requests are sent by clients
that are far from the private cloud and much closer to the public cloud,
it might be beneficial to switch to the Dog or Peacock mode.
\protocol might also plan to switch back to the Lion mode,
e.g., when the load on the private cloud is reduced.
To change the mode, the protocol also has to change the view, because the primary
and the set of participant replicas might be different in different modes.
Therefore, to handle a mode change, the protocol first performs a
view change, and then the primary of the new view in the new mode
starts to process new requests.

For the switch to happen
a {\em trusted} replica $s$ multicasts a
$\langle\text{\scriptsize \MCHANGE}, v+1, \pi' \rangle_{\sigma_s}$ to
all the replicas where $\pi'$ is the new mode of the protocol,
i.e., Lion, Dog, or Peacock.
When the protocol wants to switch to the Lion or Dog mode, replica $s$ is the primary of view $v+1$, and
when it switches to the Peacock mode, replica $s$ is the transferer of view $v+1$.

\subsection{Discussion}\label{sec:disc}

In this section, we compare the different modes of \protocol with three well-known protocols:
the crash fault-tolerant protocol Paxos \cite{lamport2001paxos},
the Byzantine fault-tolerant protocol PBFT \cite{castro1999practical}, and
the hybrid fault-tolerant protocol UpRight \cite{clement2009upright}.

We consider four parameters in this comparison:
(1) the number of communication phases,
(2) the number of message exchanges,
(3) the receiving network size, and
(4) the quorum size.
The results are reported in Table~\ref{tbl:comp}.

The knowledge of where a crash or a malicious failure may occur and thus
choosing a trusted primary
simply reduces one phase of communication.
In fact, in PBFT,
the {\sf prepare} phase is needed only to make sure that non-faulty replicas receive matching 
{\sf pre-prepare} messages from the primary.
In contrast, in the Lion and Dog modes of \protocol,
since the primary is a trusted node, replicas receive the same message
from the primary, thus there is no need for that phase of communication and
the requests, similar to Paxos, are processed in two phases
(while in contrast to Paxos malicious failures can occur in the public cloud).
In comparison to Upright,
although Upright processes the requests in two phases,
it utilizes the speculative execution technique introduced by Zyzzyva \cite{kotla2007zyzzyva}
which becomes costly in the presence of failures.
Note that the speculative execution technique can easily be applied to \protocol as well.

The number of message exchanges in the Lion mode is similar to Paxos and is linear
in terms of the total number of replicas.
In the Dog mode, the number of messages is quadratic,
however it is still much less than PBFT
(since it has one phase of $n$-to-$n$ communication instead of two).
Upright also has a quadratic number of messages.

The Lion mode, similar to Upright, needs $3m+2c+1$ nodes to receive a client request.
In the Dog mode, however, only the trusted primary and $3m+1$ nodes from the public cloud participate in each phase.
Since the Peacock mode utilizes PBFT, the number of phases and message exchanges are the same as PBFT.
However, since the primary is in the public cloud, communicating with the private cloud has no advantage,
thus it proceeds with $3m+1$ nodes instead of $3m+2c+1$ as in the Lion mode and UpRight.

{\footnotesize
\begin{table}[t]
\caption{Comparison of fault-tolerant protocols}
\label{tbl:comp}
\begin{center}
\setlength\tabcolsep{1.5pt}
\begin{tabular}{ |c|c|c|c|c|c| }
  \hline
  Protocol & phases & messages & Receiving Network & Quorum size\\
 \hline
  Lion & $2$ & $\calo(n)$ & $3m{+}2c{+}1$ & $2m{+}c{+}1$\\
 \hline
  Dog & $2$ & $\calo(n^2)$ & $3m{+}1$ & $2m{+}1$\\
 \hline
  Peacock & $3$ & $\calo(n^2)$ & $3m{+}1$ & $2m{+}1$\\
 \hline
  Paxos & $2$ & $\calo(n)$ & $2f{+}1$ & $f{+}1$\\
 \hline
  PBFT & $3$ & $\calo(n^2)$ & $3f{+}1$ & $2f{+}1$\\
 \hline
  UpRight & $2$ & $\calo(n^2)$ & $3m{+}2c{+}1$ & $2m{+}c{+}1$\\
 \hline
\end{tabular}
\end{center}
\end{table}
}

\section{Performance Evaluation}\label{sec:eval}

\begin{figure*}[t!]
\begin{minipage}{.23\textwidth}\centering
\begin{tikzpicture}[scale=0.49]
\begin{axis}[
    xlabel={Throughput [Kreqs/sec]},
    ylabel={Latency [ms]},
    xmin=0, xmax=24,
    ymin=0, ymax=9,
    xtick={0,4,8,12,16,20},
    ytick={0,2,4,6,8},
    legend pos=north west,
    ymajorgrids=true,
    grid style=dashed,
]

\addplot[
    color=magenta,
    mark=otimes,
    mark size=4pt,
    ]
    coordinates {
    (0.892,1.123)(5.457,1.477)(9.756,1.802)(13.377,2.039)(15.278,2.172)(16.051,2.400)(17.316,4.034)(17.500,8.321)};
    
\addplot[
    color=black,
    mark=diamond,
    mark size=4pt,
    ]
    coordinates {
    (0.911,1.110)(5.603,1.401)(9.756,1.702)(13.304,1.934)(15.204,2.102)(15.992,2.105)(17.602,3.594)(18.010,7.846)};
    
\addplot[
    color=violet,
    mark=*,
    mark size=4pt,
    ]
    coordinates {
    (0.952,1.090)(5.702,1.307)(9.756,1.502)(13.402,1.739)(14.920,1.873)(16.551,2.110)(18.524,2.734)(18.950,6.421)};

\addplot[
    color=red,
    mark=o,    
    mark size=4pt,
    ]
    coordinates {
    (0.962,1.002)(5.930,1.078)(12.320,1.188)(15.110,1.311)(17.930,1.410)(20.002,2.309)(20.350,5.903)};

\addplot[
    color=green,
    mark=square,
    mark size=4pt,
    ]
    coordinates {
    (0.972,1.03)(6.119,1.021)(11.730,1.292)(17.016,1.230)(20.820,1.534)(21.230,2.312)(21.401,6.732)};

\addplot[
    color=blue,
    mark=triangle,
    mark size=4pt,
    ]
    coordinates {
    (0.982,0.991)(6.349,1.021)(11.976,1.228)(17.316,1.290)(21.164,1.334)(22.099,1.989)(22.102,5.989)};

\addlegendentry{BFT}
\addlegendentry{S-UpRight}
\addlegendentry{Peacock}
\addlegendentry{Dog}
\addlegendentry{Lion}
\addlegendentry{CFT}

\end{axis}
\end{tikzpicture}
(a) $f=2$ ($c=1$, $m=1$) \newline{\scriptsize $N$: \protocol, S-UpRight=$6$, CFT=$5$, BFT=$7$}
\label{fig:c1m1}
\end{minipage}\hfill
\begin{minipage}{.23\textwidth} \centering
\begin{tikzpicture}[scale=0.49]
\begin{axis}[
    xlabel={Throughput[Kreqs/sec]},
    ylabel={Latency [ms]},
    xmin=0, xmax=15,
    ymin=0, ymax=9,
    xtick={0,3,6,9,12},
    ytick={0,2,4,6,8},
    legend pos=north west,
    ymajorgrids=true,
    grid style=dashed,
]

\addplot[
    color=magenta,
    mark=otimes,
    mark size=4pt,
    ]
    coordinates {
    (0.840,1.302)(4.819,1.560)(8.908,2.274)(9.416,2.980)(10.416,7.443)};
    
\addplot[
    color=black,
    mark=diamond,
    mark size=4pt,
    ]
    coordinates {
    (0.862,1.290)(4.803,1.501)(9.117,2.252)(9.912,2.602)(10.770,6.613)};
    
\addplot[
    color=violet,
    mark=*,
    mark size=4pt,
    ]
    coordinates {
    (0.839,1.288)(4.811,1.463)(8.732,2.002)(11.630,2.549)(11.934,6.402)};

\addplot[
    color=red,
    mark=o,
    mark size=4pt,
    ]
    coordinates {
    (0.872,1.232)(9.109,1.602)(13.219,2.601)(13.481,5.601)};

\addplot[
    color=green,
    mark=square,
    mark size=4pt,
    ]
    coordinates {
    (0.871,1.102)(9.219,1.542)(13.510,2.431)(13.704,5.101)};
    
\addplot[
    color=blue,
    mark=triangle,
    mark size=4pt,
    ]
    coordinates {
    (0.921,1.002)(9.526,1.668)(10.610,1.865)(14.345,2.273)(14.738,5.076)};

\addlegendentry{BFT}
\addlegendentry{S-UpRight}
\addlegendentry{Peacock}
\addlegendentry{Dog}
\addlegendentry{Lion}
\addlegendentry{CFT}

\end{axis}
\end{tikzpicture}
(b) $f=4$ ($c=2$, $m=2$) \newline{\scriptsize $N$: \protocol, S-UpRight=$11$, CFT=$9$, BFT=$13$} 
\label{fig:c2m2}
\end{minipage}\hfill
\begin{minipage}{.23\textwidth} \centering
\begin{tikzpicture}[scale=0.49]
\begin{axis}[
    xlabel={Throughput [Kreqs/sec]},
    ylabel={Latency [ms]},
    xmin=0, xmax=15,
    ymin=0, ymax=9,
    xtick={0,3,6,9,12},
    ytick={0,2,4,6,8},
    legend pos=north west,
    ymajorgrids=true,
    grid style=dashed,
]

\addplot[
    color=magenta,
    mark=otimes,
    mark size=4pt,
    ]
    coordinates {
    (0.840,1.302)(4.819,1.560)(8.908,2.274)(9.416,2.980)(10.416,7.443)};
    
\addplot[
    color=black,
    mark=diamond,
    mark size=4pt,
    ]
    coordinates {
    (0.862,1.284)(4.833,1.541)(8.975,2.203)(9.612,2.802)(10.511,7.102)};
    
\addplot[
    color=violet,
    mark=*,
    mark size=4pt,
    ]
    coordinates {
    (0.839,1.288)(4.948,1.532)(8.732,2.102)(10.002,2.731)(10.832,6.102)};

\addplot[
    color=red,
    mark=o,
    mark size=4pt,
    ]
    coordinates {
    (0.872,1.232)(5.021,1.321)(8.923,1.902)(10.489,2.434)(11.745,6.601)};

\addplot[
    color=green,
    mark=square,
    mark size=4pt,
    ]
    coordinates {
    (0.920,1.102)(8.921,1.613)(13.002,2.543)(13.248,5.31)};

\addplot[
    color=blue,
    mark=triangle,
    mark size=4pt,
    ]
    coordinates {
    (0.921,1.002)(9.526,1.668)(10.610,1.865)(14.345,2.273)(14.738,5.076)};
    
\addlegendentry{BFT}
\addlegendentry{S-UpRight}
\addlegendentry{Peacock}
\addlegendentry{Dog}
\addlegendentry{Lion}
\addlegendentry{CFT}

\end{axis}
\end{tikzpicture}
(c) $f=4$ ($c=1$, $m=3$) \newline{\scriptsize $N$: \protocol, S-UpRight=$12$, CFT=$9$, BFT=$13$}
\label{fig:c1m3}
\end{minipage}\hfill
\begin{minipage}{.23\textwidth} \centering
\begin{tikzpicture}[scale=0.49]
\begin{axis}[
    xlabel={Throughput [Kreqs/sec]},
    ylabel={Latency [ms]},
    xmin=0, xmax=21,
    ymin=0, ymax=9,
    xtick={0,4,8,12,16,20},
    ytick={0,2,4,6,8},
    legend pos=north west,
    ymajorgrids=true,
    grid style=dashed,
]
 
\addplot[
    color=magenta,
    mark=otimes,
    mark size=4pt,
    ]
    coordinates {
    (0.840,1.302)(4.819,1.560)(8.908,2.274)(9.416,2.980)(10.416,7.443)};
    
\addplot[
    color=black,
    mark=diamond,
    mark size=4pt,
    ]
    coordinates {
    (0.911,1.170)(5.103,1.401)(9.943,2.152)(10.902,2.543)(11.701,6.413)};
    
\addplot[
    color=violet,
    mark=*,
    mark size=4pt,
    ]
    coordinates {
    (0.952,1.090)(5.702,1.307)(9.756,1.502)(13.402,1.739)(14.920,1.873)(16.551,2.110)(18.524,2.734)(18.950,6.421)};

\addplot[
    color=red,
    mark=o,
    mark size=4pt,
    ]
    coordinates {
    (0.962,1.002)(5.930,1.078)(12.320,1.188)(15.110,1.311)(17.930,1.410)(20.002,2.309)(20.350,5.903)};

\addplot[
    color=green,
    mark=square,
    mark size=4pt,
    ]
    coordinates {
    (0.903,1.0432)(9.413,1.473)(14.004,2.378)(14.304,5.003)};

\addplot[
    color=blue,
    mark=triangle,
    mark size=4pt,
    ]
    coordinates {
    (0.921,1.002)(9.526,1.668)(10.610,1.865)(14.345,2.273)(14.738,5.076)};
    
\addlegendentry{BFT}
\addlegendentry{S-UpRight}
\addlegendentry{Peacock}
\addlegendentry{Dog}
\addlegendentry{Lion}
\addlegendentry{CFT}

\end{axis}
\end{tikzpicture}
(d) $f=4$ ($c=3$, $m=1$) \newline{\scriptsize $N$: \protocol, S-UpRight=$10$, CFT=$9$, BFT=$13$}
\label{fig:c3m1}
\end{minipage}
\caption{Throughput/Latency measurement by increasing the number of failures}
  \label{fig:faultscale}
\end{figure*}
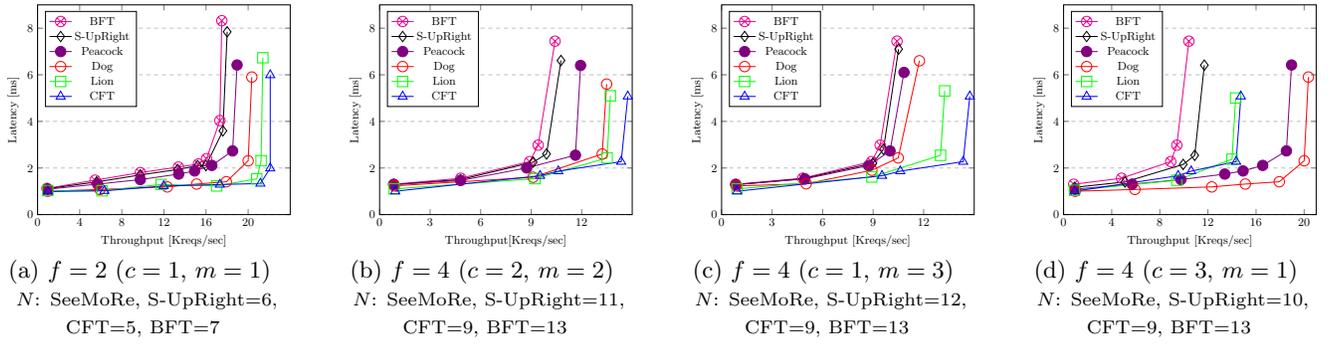

This section evaluates the performance of the \protocol protocol.
\protocol is implemented by adapting the BFT-SMaRt
library \cite{sousa2013state}.
We mainly reuse the communication layer of BFT-SMaRt
but implement our agreement and view change routines for the different modes
of the protocol.

We first, show how the protocol tolerates
different number of failures (both crash and malicious).
Next, we evaluate the performance of the protocol
in a no failure setting by varying the
number of clients (requests) and using different micro-benchmarks, and finally,
evaluate the impact of the failure of the primary node (view change)
on the performance of \protocol.

In each experiment, we compare different modes of \protocol with
an asynchronous crash fault-tolerant (CFT) protocol,
an asynchronous Byzantine fault-tolerant (BFT) protocol, and
a simplified version of the asynchronous hybrid fault-tolerant protocol UpRight \cite{clement2009upright}
(we call it {\em S-UpRight}).
For both CFT and BFT we use the original BFT-SMaRt codebase
(the optimized implementations of Paxos \cite{lamport2001paxos} and PBFT \cite{castro1999practical}).
UpRight has two aspects: first, the hybrid model that tolerates both crash and malicious failures
(in a network of size $3m+2c+1$), and
second, the protocol that combines a set of techniques such as
speculative execution and separation of ordering and execution.
For the S-UpRight protocol, we use the UpRight hybrid model
since this part of the UpRight is relevant to \protocol, however,
to ensure a fair comparison with other protocols and
since all other protocols use the pessimistic approach,
we use a PBFT-like protocol (i.e., PBFT protocol with less number of nodes)
instead of the UpRight protocol.
Note that, as mentioned before, both the speculative execution and
separation of ordering from execution techniques can be integrated into \protocol as well.

The experiments were conducted on the Amazon EC2 platform.
Each VM is Compute Optimized c4.2xlarge instances with 8 vCPUs and 15GB RAM,
Intel Xeon E5-2666 v3 processor clocked at 3.50 GHz.
In the experiments, both the public and private clouds are located in the same data center i.e., AWS US West Region.

In each experiment, we vary the number of requests sent by all the clients per second from $10^3$ to $10^6$
(by increasing the number of clients running on a single VM)
and measure the end-to-end throughput ($x$ axis) and latency ($y$ axis) of the system.
Each client waits for the reply before sending a subsequent request.

\subsection{Fault-Tolerance Scalability}

In the first set of experiments, we evaluate the performance of \protocol
when configured to tolerate different number of failures ($f$).
We consider the $0/0$ micro-benchmark
(both request and reply payload sizes are close to $0$ KB)
and measure
the throughput and latency of \protocol, S-UpRight, CFT, and BFT protocols. 
Since, $f=c+m$,
we evaluate CFT and BFT to tolerate $c+m$ failures in each experiment.
In all these scenarios and for \protocol, we put $2c$ nodes in the private
and $3m+1$ nodes in the public cloud.
The results are shown in Fig.~\ref{fig:faultscale}(a)-(d).

In the first scenario, when $f=2$ ($c=m=1$),
the network size of the different protocols is close to each other
(BFT requires $7$, \protocol and S-UpRight require $6$, and CFT requires $5$ nodes).
As a result,
as can be seen in Fig.~\ref{fig:faultscale}(a),
the performance of the Lion mode becomes very close to CFT
($8\%$ difference in their peak throughput).
Similarly, the performances of S-UpRight and BFT are close to each other
($4\%$ difference in their peak throughput).
Note that the Peacock mode shows better performance than S-UpRight (still worst than the Dog and Lion modes) because
in the Peacock mode, \protocol relies only on the public cloud which consists of only $4$ nodes.
In addition, while in comparison to the Lion mode, both the Peacock and Dog modes need less number of nodes,
the Lion mode has better performance because it needs less number of phases and message exchanges.

In the next three scenarios, the network tolerates the same number of failures ($f=4$), as a result,
the performance of BFT and CFT does not change from one scenario to another.
However, since the number of crash and malicious failures are varied,
the network size of \protocol and S-UpRight changes. Hence,
they show different performance in different scenarios.

When both $m$ and $c$ increase to $2$ (Fig.~\ref{fig:faultscale}(b)),
The Dog mode shows similar performance to the Lion mode.
This is the result of the trade off between the quorum size and the message complexity;
Only $5$ nodes ($2m+1$) participate in the Dog mode which requires $\calo(n^2)$ number of messages
whereas the quorum size of the Lion mode is $7$ ($2m+1c+1$) but it requires $\calo(n)$ messages (see Table~\ref{tbl:comp}).
In addition, since \protocol in the Peacock mode communicates with only $7$ nodes,
it shows much better performance than BFT ($24\%$ difference in their peak throughput) and
even S-UpRight ($18\%$ difference in their peak throughput).

By increasing the number of tolerated malicious failures to $3$
while reducing the number of tolerated crash failures back to $1$ (Fig.~\ref{fig:faultscale}(c)),
the network size of \protocol becomes closer to the BFT network size.
As a result, CFT shows better performance ($12\%$ difference in its peak throughput) than the Lion mode and
also the performance of \protocol in the Peacock and Dog modes, which communicate with $10$ nodes in the public cloud,
becomes closer to S-UpRight (with $12$ nodes) and BFT (with $13$ nodes).

On the other hand, increasing the number of tolerated crash failures to $3$
while maintaining the number of malicious failures to $1$ (Fig.~\ref{fig:faultscale}(d))
results in a network size close to CFT.
In this setting, the performance of the Dog and Peacock modes become better than both the Lion mode and CFT. 
This is expected because the Dog mode processes a request in the public cloud which needs only $4$ replicas (since $m=1$)
but with the same number of phases as the Lion mode. Similarly, although the Peacock mode processes requests in three phases,
since it needs fewer servers to proceed, its performance is better than the Lion mode and CFT.
In fact, since the number of malicious failures in this scenario is the same as the first scenario,
both the Dog and Peacock modes show the same performance as the first scenario (Fig.~\ref{fig:faultscale}(a)).

\subsection{Changing Payload Size}

\begin{figure}[t!]
\begin{minipage}{.23\textwidth}\centering
\begin{tikzpicture}[scale=0.5]
\begin{axis}[
    xlabel={Throughput [kreqs/sec]},
    ylabel={Latency [ms]},
    xmin=0, xmax=23,
    ymin=0, ymax=13.000,
    xtick={0,5,10,15,20},
    ytick={0,2,4,6,8,10,12},
    legend pos=north west,
    ymajorgrids=true,
    grid style=dashed,
]

\addplot[
    color=magenta,
    mark=otimes,
    mark size=3pt,
    ]
    coordinates {
    (5.095,1.767)(9.661,1.999)(13.745,3.110)(15.873,5.131)(15.902,11.100)};
    
\addplot[
    color=black,
    mark=diamond,
    mark size=3pt,
    ]
    coordinates {
    (5.195,1.701)(9.892,1.911)(14.013,3.043)(16.248,5.102)(16.449,10.803)};
    
\addplot[
    color=violet,
    mark=*,
    mark size=3pt,
    ]
    coordinates {
    (6.320,1.467)(9.461,1.599)(13.630,2.719)(15.027,3.671)(17.200,5.291)(17.400,10.435)};

\addplot[
    color=red,
    mark=o,
    mark size=3pt,
    ]
    coordinates {
    (5.470,1.201)(11.328,1.402)(13.809,2.410)(15.830,3.501)(17.131,4.593)(19.101,6.293)(19.530,10.092)};

\addplot[
    color=green,
    mark=square,
    mark size=3pt,
    ]
    coordinates {
    (5.913,1.104)(10.973,1.299)(14.201,2.210)(16.091,3.131)(18.020,4.100)(20.543,6.01)(20.900,9.012)};

\addplot[
    color=blue,
    mark=triangle,
    mark size=3pt,
    ]
    coordinates {
    (6.106,1.050)(11.428,1.270)(13.698,2.013)(16.736,3.100)(19.400,4.333)(21.356,6.013)(21.540,10.012)};

\addlegendentry{BFT}
\addlegendentry{S-UpRight}
\addlegendentry{Peacock}
\addlegendentry{Dog}
\addlegendentry{Lion}
\addlegendentry{CFT}

\end{axis}
\end{tikzpicture}
(a) Benchmark 0/4
\label{fig:bench04}
\end{minipage}\hfill
\begin{minipage}{.23\textwidth} \centering
\begin{tikzpicture}[scale=0.5]
\begin{axis}[
    xlabel={Throughput [kreqs/sec]},
    ylabel={Latency [ms]},
    xmin=0, xmax=15,
    ymin=0, ymax=13.000,
    xtick={0,3,6,9,12,15},
    ytick={0,2,4,6,8,10,12},
    legend pos=north west,
    ymajorgrids=true,
    grid style=dashed,
]

\addplot[
    color=magenta,
    mark=otimes,
    mark size=3pt,
    ]
    coordinates {
    (4.474,2.299)(7.843,3.029)(9.195,4.407)(11.111,6.986)(11.378,11.811)};
    
\addplot[
    color=black,
    mark=diamond,
    mark size=3pt,
    ]
    coordinates {
    (4.574,2.173)(8.031,3.041)(9.243,4.007)(10.875,5.659)(11.738,6.934)(12.081,10.942)};
    
\addplot[
    color=violet,
    mark=*,
    mark size=3pt,
    ]
    coordinates {
    (4.701,2.043)(8.201,3.055)(8.870,3.483)(10.923,4.764)(12.084,6.188)(13.102,11.418)};

\addplot[
    color=red,
    mark=o,
    mark size=3pt,
    ]
    coordinates {
    (5.501,1.843)(8.420,2.505)(10.800,3.401)(12.645,5.964)(13.701,10.001)};

\addplot[
    color=green,
    mark=square,
    mark size=3pt,
    ]
    coordinates {
    (5.209,1.544)(8.531,2.253)(10.983,3.191)(12.940,5.322)(13.773,7.419)(14.001,10.409)};

\addplot[
    color=blue,
    mark=triangle,
    mark size=3pt,
    ]
    coordinates {
    (5.319,1.643)(8.869,2.355)(11.267,3.283)(13.245,5.464)(14.084,7.588)(14.511,11.518)};

\addlegendentry{BFT}
\addlegendentry{S-UpRight}
\addlegendentry{Peacock}
\addlegendentry{Dog}
\addlegendentry{Lion}
\addlegendentry{CFT}

\end{axis}
\label{fig:bench400}
\end{tikzpicture}
(b) Benchmark 4/0
\label{fig:bench40}
\end{minipage}
\caption{Throughput/Latency for $c=1$ and $m=1$}
  \label{fig:benchmarks}
\end{figure}
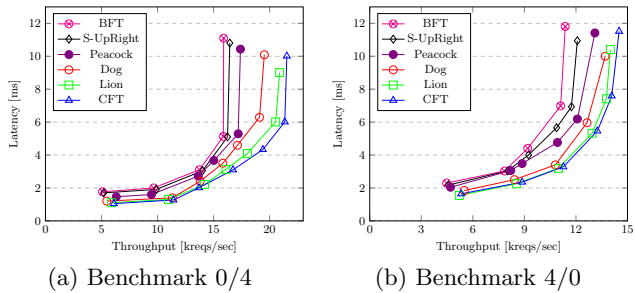

We now repeat the base case scenario ($c{=}m{=}1$) of the previous experiments (Fig.~\ref{fig:faultscale}(a))
using two micro-benchmarks $0/4$, $4/0$ to show how request and reply sizes affect the performance of
different protocol.
Figs.~\ref{fig:benchmarks}(a) and \ref{fig:benchmarks}(b) show
the throughput and latency for $0/4$ and $4/0$ micro-benchmarks respectively.
Since the Lion and Dog modes need less communication phases and message exchanges,
their performance is close to CFT, e.g.,
for latency equal to $4$ ms, the throughput of the Lion and Dog modes is $10\%$ and $17\%$ less than CFT respectively.
Similarly, the Peacock mode and S-UpRight are close to BFT, e.g.,
with $4$ ms latency, the throughput of the Peacock mode is the same as BFT.
Note that due to the overhead of request transmission between the replicas,
the request size affects the performance of all protocols more than the reply size.

\subsection{Performance During View Change}
\begin{figure}[t!]
\begin{tikzpicture}
\tikzstyle{every node}=[font=\tiny]
\begin{axis}[
     xlabel={Timeline [ms]},
     ylabel={Throughput [Kreq/s]},
     y=0.46cm/3,
    x=0.065cm,
     xmin=0, xmax=110,
     ymin=0, ymax=22,
     xtick={0,20,40,60,80,100},
     ytick={0,4,8,12,16,20},
legend style={at={(axis cs:70,000)},anchor=south west}, 
    ymajorgrids=true,
    grid style=dashed,
]

\addplot[
    color=magenta,
    mark=otimes,
    ]
    coordinates {
    (0,12.340)(5,12.350)(10,12.330)(15,12.340)(20,12.340)(25,12.430)(30,12.410)(32,0)(61,0)(61,12.420)(60,12.624)(65,12.601)(70,12.231)(75,12.340)(80,12.340)(85,12.340)(90,12.613)(95,12.091)};
    
\addplot[
    color=black,
    mark=diamond,
    ]
    coordinates {
    (0,13.110)(5,13.253)(10,13.439)(15,13.492)(20,13.293)(25,13.502)(30,13.478)(32,0)(59,0)(59,13.510)(60,13.484)(65,13.801)(70,13.931)(75,13.804)(80,13.503)(85,13.593)(90,13.464)(95,13.618)};
    
\addplot[
    color=violet,
    mark=*,
    ]
    coordinates {
    (0,15.021)(5,15.429)(10,15.721)(15,15.493)(20,15.502)(25,15.803)(30,16.002)(32,0)(56,0)(56,15.792)(55,15.731)(60,15.993)(65,15.982)(70,16.021)(75,15.992)(80,15.892)(85,16.102)(90,16.002)(95,15.992)};
\addplot[
    color=red,
    mark=o,
    ]
    coordinates {
    (0,17.231)(5,17.492)(10,17.692)(15,17.411)(20,17.413)(25,17.801)(30,17.973)(31,0)(51,0)(51,17.813)(55,17.702)(60,17.992)(65,17.974)(70,17.982)(75,17.973)(80,17.934)(85,17.874)(90,17.981)(95,17.940)};

\addplot[
    color=green,
    mark=square,
    ]
    coordinates {
    (0,19.189)(5,19.436)(10,19.623)(15,19.375)(20,19.365)(25,19.721)(30,19.912)(32,0)(48,0)(48,19.724)(50,19.724)(55,19.634)(60,19.913)(65,19.913)(70,19.913)(75,19.913)(80,19.871)(85,19.824)(90,19.913)(95,19.913)};

\addlegendentry{BFT}
\addlegendentry{S-Upright}
\addlegendentry{Peacock}
\addlegendentry{Dog}
\addlegendentry{Lion}

\end{axis}
\end{tikzpicture}
\caption{Performance during view change}
  \label{fig:faultprimary}
\end{figure}
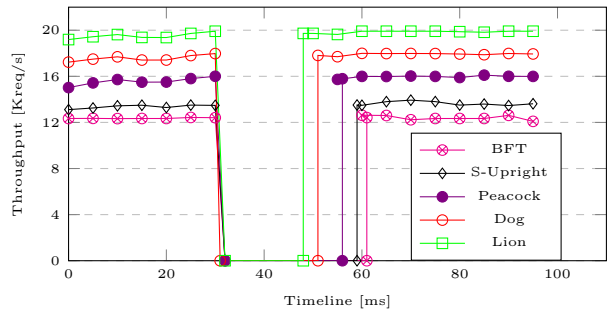

Finally, we evaluate the impact of view change on the performance of
\protocol.
We trigger a primary failure shortly before
the end of a checkpoint period to evaluate the worst-case
overhead that can be caused by a failure.
We consider the base case scenario ($c=m=1$) with a total network of $N=6$ nodes (for \protocol),
where $2$ nodes are in the private cloud and $4$ in the public cloud.
The experiment was run with micro-benchmark $0/0$ and with a checkpoint period of $10000$ request i.e., a checkpoint is taken every $10000$ requests.
Fig.~\ref{fig:faultprimary} shows the behavior of \protocol, S-UpRight and BFT where the $y$-axis is throughput
and the $x$-axis is a timeline with a failure injected around time $30$.
As can be seen, the protocols behave as expected until the
crash is triggered. 
This crash and the view change routine cause the protocols to
be temporarily out of service
(in particular, $15$ millisecond in the Lion mode,
$20$ millisecond in the Dog mode, and
$24$ millisecond in the Peacock mode).
However, when the view change is complete,
the throughput increases to the original level for each protocol.
As can be seen, BFT takes twice as much time as the Lion mode to revive and continue to process the requests.
The Peacock mode also recovers faster than S-UpRight and BFT due to its use of transferers.

Overall, the evaluation results for a network that tolerates $f=m+c$ failures where
$m$ and $c$ are the number of malicious and crash failures respectively, can be summarized as follow.
First, when $c$ is equal or less than $m$ (for small $c$ and $m$),
the performance of \protocol in the Lion mode is very close to the crash fault-tolerant protocol Paxos due to
the required number of phases and message exchanges in the Lion mode.
In addition, when $c$ is larger than $m$,
\protocol in both the Dog and Peacock modes demonstrates better performance than the Lion mode and even Paxos
since in both modes, \protocol relies completely on the public cloud to process the requests.
Furthermore, all three modes of \protocol show better performance than the hybrid protocol S-UpRight since \protocol is aware of
where the crash faults may occur and where the malicious faults can occur.
Finally, all three modes of \protocol have better performance than BFT since they reduce
the number of communication phases, messages exchanged and required nodes.

\newpage
\section{Conclusions}\label{sec:conc}
In this paper, we proposed \protocol, a hybrid state machine replication protocol
to tolerate both crash and malicious failures in a public/private cloud environment.
\protocol is targeted to be used by smaller enterprises that own a small set of servers
and intend to rent servers from public cloud providers. Such an
enterprise can highly benefit
from \protocol, as the protocol distinguishes between crash failures that could occur 
within the trusted private cloud and malicious failures that could only occur in the public cloud. 
\protocol can execute in any one of three modes, Lion, Dog, and Peacock, and can dynamically switch
among these modes.
The Lion and Dog modes of \protocol require less communication phases and message exchanges while
the Peacock mode is useful for a heavily loaded private cloud or
when there is a large network distance between the two clouds.

Our evaluations show that the performance of Lion and Dog modes is close to Paxos
while in contrast to Paxos, which only tolerates crash failures, malicious failures can occur.
In the Peacock mode, since the primary is in the public cloud,
its performance is similar to PBFT with $m$ failures.
However, in comparison to UpRight, which requires quorums of size $2m+c+1$,
Peacock needs quorums of size $2m+1$, and hence is more efficient.

\balance

\bibliographystyle{abbrv}
\bibliography{main}

\end{document}